\pdfoutput=1

\documentclass[final,5p,times,twocolumn,12pt]{elsarticle}

\usepackage{amssymb}
\usepackage{hyperref}
\usepackage{cuted}
\usepackage{breakurl,epstopdf,xcolor}
\usepackage{dcolumn}

\biboptions{sort&compress}

	\usepackage{fancyheadings}
\renewcommand{\thispagestyle}[1]{} 

\begin{document}

\begin{frontmatter}



\title{Magnetocaloric properties of V6 molecular magnet}


\author{P. Kowalewska\fnref{label1}}

\fntext[label2]{\href{https://orcid.org/0000-0003-4254-2492}{\includegraphics[height=12pt]{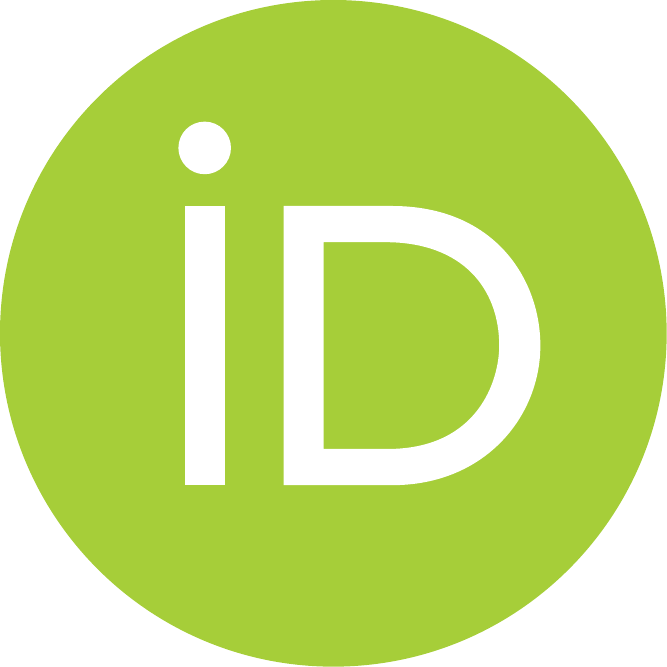}$\,\,$https://orcid.org/0000-0003-4254-2492}\\$\,$\\$\,$\\}

\author{K. Sza\l{}owski\corref{cor1}\fnref{label2}}
\ead{karol.szalowski@uni.lodz.pl}

\fntext[label2]{\href{https://orcid.org/0000-0002-3204-1849}{\includegraphics[height=12pt]{orcid.eps}$\,\,$https://orcid.org/0000-0002-3204-1849}}

\address{University of \L\'{o}d\'{z}, Faculty of Physics and Applied Informatics, Department of Solid State Physics,\\ulica Pomorska 149/153, 90-236 \L\'{o}d\'{z}, Poland}

\cortext[cor1]{Corresponding author}

\begin{abstract}
The paper presents a theoretical study of magnetocaloric properties of polyoxovanadate molecular magnet V6 containing 6 vanadium ions carrying quantum spins $S = 1/2$. The characteristic property of such structure is the presence of two weakly interacting spin triangles with all-antiferromagnetic couplings. The properties of the system are described using the exact numerical diagonalization approach applied to the quantum Heisenberg model and utilizing a field ensemble formalism. The dependence of the magnetic entropy and magnetic specific heat on the temperature and external magnetic field is calculated and extensively discussed. The magnetocaloric properties are quantified by isothermal entropy change and entropy derivative over the magnetic field. An interesting behaviour of isothermal entropy change is found, with high degree of tunability of the magnetocaloric effect with the initial and final magnetic field values. 
\begin{center}
\includegraphics[width=0.60\textwidth]{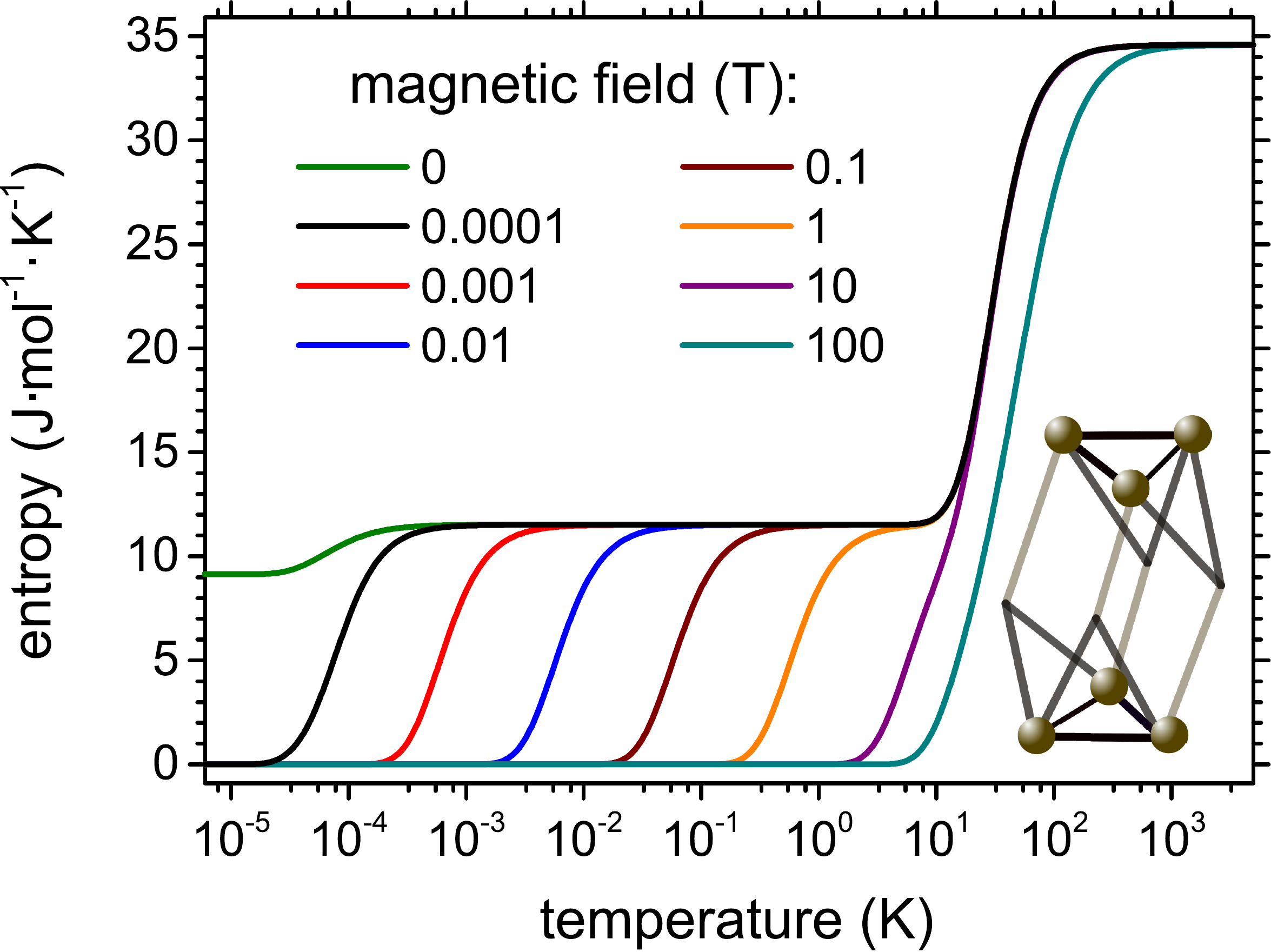}
\end{center}
\end{abstract}


\end{frontmatter}



\section{Introduction}
\label{intro}
Low-dimensional magnetic systems attract considerable attention due to their unique properties. One of their experimental realizations is based on a wide class of molecular magnets \cite{Pinkowicz2011,Sieklucka2017,Blundell2004}, within which single molecule magnets provide fascinating examples of zero-dimensional magnetic clusters \cite{Gatteschi2006,Friedman2010}. Among the huge variety of molecular magnetic materials, systems containing a triangle as a fundamental unit constitute a highly interesting category \cite{Ponomaryov2015,Furukawa2016}. This is, mainly, due to the fact that antiferromagnetically coupled triangle is an archetypical example of magnetically frustrated system \cite{Diep2013}. The mentioned class of molecular magnets includes those based on Mn ions \cite{Hu2015} or Cu ions \cite{Choi2008,Ponomaryov2015, Hao2013,Nath2013,Iida2011}. A distinct group within this class consists of various polyoxovanadates containing $S = 1/2$ spins originating from vanadium ions – gathering such examples as V3 \cite{Muller1998, Iida2009, Belinsky2014, Belinsky2016, Belinsky2016a}, V6 \cite{Corti2011,Luban2002,Rousochatzakis2005,Jung2002,Haraldsen2009}, V12 \cite{Barbour2006} and particularly frequently investigated V15 \cite{Fu2015,Chiorescu2000,Chaboussant2002,Tarantul2010, Tarantul2011,Furukawa2005,Furukawa2007,Salman2009,Kostyuchenko2008, Konstantinidis2002, Wernsdorfer2004,Vongtragool2003,Sakon2005,Gysler2014,Procissi2006, Tsukerblat2018, Furukawa2007a, Popov2004} clusters. 

Magnetocaloric effect, consisting in dependence of entropy on the applied magnetic field, bears high potential for modern applications and attracts rising attention within the community of condensed matter physicists \cite{Franco2018,deOliveira2010a}. Therefore, one of the highly interesting challenges for physics of molecular magnets is characterization of their magnetocaloric properties and constant search for optimization of the magnetocaloric performance \cite{Sessoli2012,Evangelisti2006,Zheng2014,Garlatti2013,
Liu2016g,Evangelisti2014,Evangelisti2010,Liu2014,Sharples2013}. Experimental studies focused on such qualities have been reported for some classes of systems, to mention Fe-based systems \cite{Evangelisti2005,Torres2000} or Mn-based ones \cite{Balanda2016,Torres2000,Evangelisti2009} as well as molecules incorporating Gd \cite{Pineda2016}. The experimental results are supplemented with some theoretical insights involving calculations on the grounds of density functional theory present in the recent literature, in particular for V15 \cite{Kortus2001,Boukhvalov2004,Kortus2001a}. Moreover, sample magnetic nanoclusters of various geometries and shapes were studied rigorously \cite{Zukovic2014a,Zukovic2015a,Zukovic2018a,Strecka2015c,
Karlova2017b,Roxburgh2018,Mohylna2019a} and particular molecular cluster systems were investigated with close to exact methods \cite{Schnack2013,Schnack2013b,Raghu2001}.

Among the mentioned vanadium-based molecular magnets, V15 attracted some attention from the magnetocaloric point of view \cite{Fu2015}. However, analogous properties of the smaller molecule, V6, have not been characterized so far. Motivated by this fact, we undertake calculations aimed at prediction of the magnetocaloric effect in V6 molecular magnet. The foundation of our calculations is the exact diagonalization of a model spin Hamiltonian followed by application of the canonical (field) ensemble formalism. The next sections of the paper present the theoretical model, the discussion of obtained numerical results and some final remarks.

\section{\label{theory}Theoretical model}

\begin{figure}[ht!]
  \begin{center}
   \includegraphics[width=0.5\columnwidth]{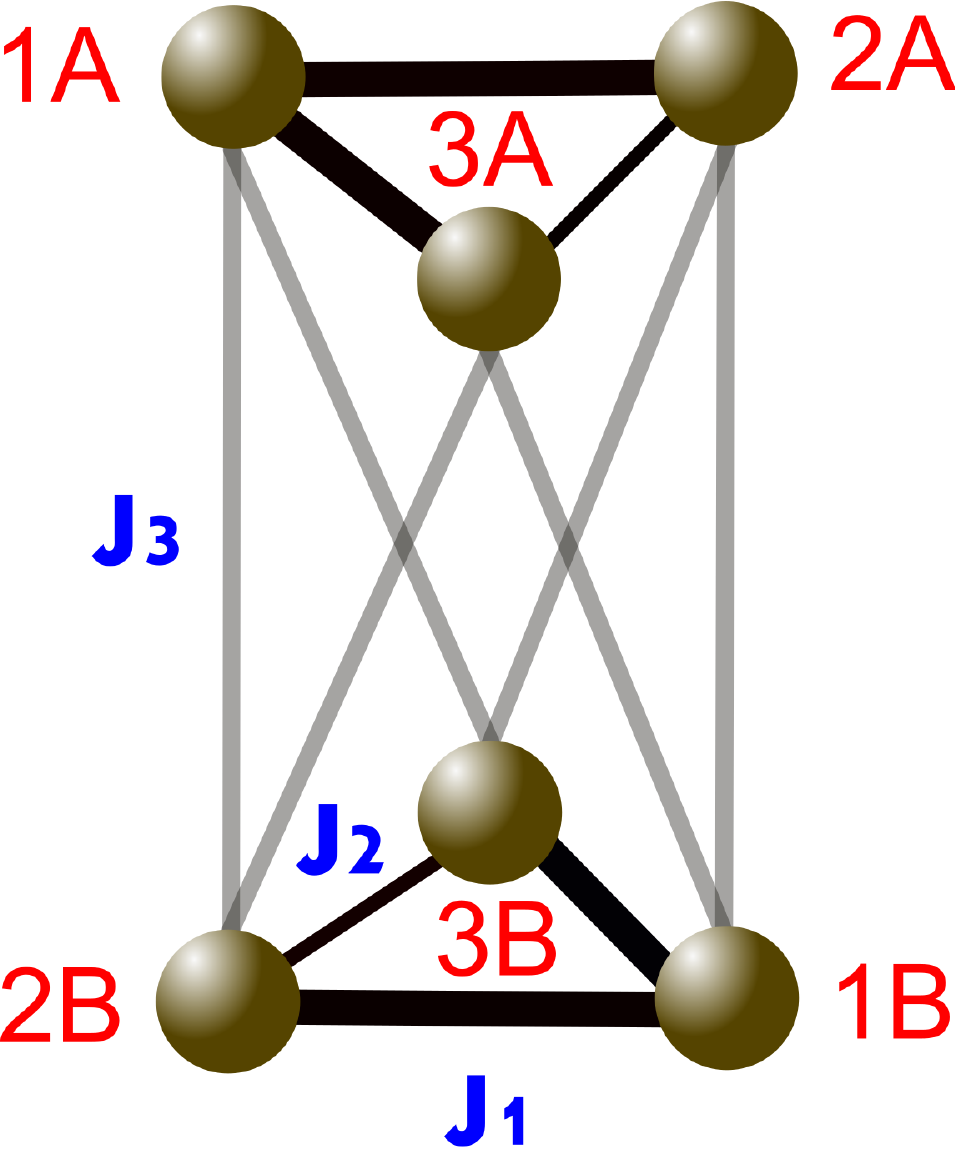}
  \end{center}
   \caption{\label{fig:figscheme} Schematic view of the V6 molecule with the positions of spins $S=1/2$, arranged in two triangles, marked with labelled circles. The solid lines correspond to the exchange interactions: intratriangle $J_1$ (thick black lines), intratriangle $J_2$ (thin black lines) and intertriangle $J_3$ (thin gray lines). }
\end{figure}

Our system of interest is a molecular magnet denoted usually by V6. In principle this refers to two polyoxovanadates: \mbox{Na$_6${H$_4$V$_6$O$_8$(PO$_4$)$_4$[(OCH$_2$)$_3$CCH$_2$OH]$_2$}$\cdot$18H$_2$O and} \mbox{(CN$_3$H$_6$)$_4$ Na$_2$ {H$_4$ V$_6$ O$_8$ (PO$_4$)$_4$ [(OCH$_2$)$_3$CCH$_2$OH]$_2$}} $\cdot$14H$_2$O \cite{Muller1998,Corti2011,Luban2002,Haraldsen2009}. However, both mentioned substances share the same substructure describing the location and interactions of spins. Namely, V6 contains 6 vanadium ions V$^{4+}$, each of them carrying a quantum spin $S = 1/2$. They are arranged into two triangles, one situated above the other (for schematic view of the magnetic ions arrangement see Fig.~\ref{fig:figscheme}). The magnetic interactions of primary importance couple the spins within each triangle (as denoted with bold black lines in Fig.~\ref{fig:figscheme}). In addition, a weak intertriangle coupling is also present in the system (marked with gray lines in Fig.~\ref{fig:figscheme}).

The system of spins in V6 cluster can be modelled with the following quantum Heisenberg Hamiltonian: 

\begin{eqnarray}\label{h}
\mathcal{H}&=&-J_1\left(\mathbf{S}_{1A}\cdot\mathbf{S}_{2A}+\mathbf{S}_{1A}\cdot\mathbf{S}_{3A}+\mathbf{S}_{1B}\cdot\mathbf{S}_{2B}\right.\nonumber\\
&+&\left.\mathbf{S}_{1B}\cdot\mathbf{S}_{3B}\right)-J_2\left(\mathbf{S}_{2A}\cdot\mathbf{S}_{3A}+\mathbf{S}_{2B}\cdot\mathbf{S}_{3B}\right)\nonumber\\
&-&J_3\left(\mathbf{S}_{1A}\cdot\mathbf{S}_{2B}+\mathbf{S}_{1A}\cdot\mathbf{S}_{3B}+\mathbf{S}_{2A}\cdot\mathbf{S}_{1B}\right.\nonumber \\&+&\left.\mathbf{S}_{2A}\cdot\mathbf{S}_{3B}+\mathbf{S}_{3A}\cdot\mathbf{S}_{1B}+\mathbf{S}_{3A}\cdot\mathbf{S}_{2B}\right)\nonumber\\
&-&g\mu_{\rm B}B\left(S^{z}_{1A}+S^{z}_{2A}+S^{z}_{3A}+S^{z}_{1B}+S^{z}_{2B}\right.\nonumber\\ &+&\left. S^{z}_{3B}\right),
\end{eqnarray} 
where the operator $\mathbf{S}_{iL}=\left(S^{x}_{iL},S^{y}_{iL},S^{z}_{iL}\right)$ is the operator of spin $S = 1/2$ localized in the triangle $L = $ A or B at $i$-th site ($i = 1,2,3$) – see Fig.~\ref{fig:figscheme} for explanation. The antiferromagnetic exchange integrals $J_1/k_{\rm B}$ = -65 K, $J_2/k_{\rm B}$ = -7 K and $J_3/k_{\rm B}$ = -0.3 K are taken from the experimental work Ref.~\cite{Rousochatzakis2005} (please note that both chemical variants of V6 actually differ a little in the values of exchange integrals – for details see Ref.~\cite{Haraldsen2009}). The fact that one of the antiferromagnetic couplings in the triangle is much weaker than the other two, $|J_2|\ll|J_1|$, is because only two out of three vanadium ions within each triangle are connected with a strong V-O-P-O-V exchange path - for details see Fig.~1 and Appendix in Ref.~\cite{Luban2002}. Note that the distances between the V ions in the triangle are nearly equal \cite{Luban2002}. The even weaker interaction $J_3$ emerges because the triangles are interlinked with four phosphate ligands - see Ref.~\cite{Luban2002} for details. The external magnetic field oriented along $z$ direction is denoted with $B$, whereas $\mu_{\rm B}$ is Bohr magneton and $g = 2$ is the gyromagnetic factor. The system Hamiltonian is represented as a matrix of the size 64$\times$64.

The thermodynamics of the system is described within the field ensemble approach (for an extensive discussion of the relations between the field ensemble and canonical ensemble for magnetic systems see Ref.~\cite{Plascak2014}). Please note that the term introducing the external magnetic field is contained in the Hamiltonian itself. The statistical sum equals $\mathcal{Z}=\sum_{k}^{}{e^{-E_k/\left(k_{\rm B}T\right)}}$, with $k_{\rm B}$ being Boltzmann constant and $T$ denoting the temperature. Its calculation is enabled by the knowledge of the complete set of eigenvalues of the system Hamiltonian $E_k$. The eigenvalues are calculated by means of exact numerical diagonalization of the Hamiltonian using Wolfram Mathematica software \cite{Mathematica}, which is used also for all the further calculations. The thermal average value of Hamiltonian $\mathcal{H}$ is equal to the magnetic free enthalpy of the system (as the external magnetic field is contained in the Hamiltonian) and is expressed by $\left\langle H\right\rangle =\left(1/\mathcal{Z}\right) \sum_{k}^{}{E_{k}\,e^{-E_k/\left(k_{\rm B}T\right)}}$. On the other hand, the magnetic Gibbs free energy is calculated directly from the statistical sum by $G=-k_{\rm B} T \ln \mathcal{Z}$. The quantity of fundamental interest in our study is the magnetic entropy of the system S and its dependence on the temperature and external magnetic field. The entropy can be calculated from the relation  $S(T,B)=\left(\left\langle H\right\rangle -G\right)/T$. Another quantity which we discuss further is the magnetic specific heat, which can be conveniently calculated with the help of the fluctuation-dissipation theorem as $c_B=\left(\left\langle H^2\right\rangle -\left\langle H\right\rangle^2 \right)/\left(k_{\rm B} T^2 \right)$.

In the paper our interest is focused on the magnetocaloric effect. One of its most crucial quantitative measures is isothermal entropy change. If the field is varied at constant temperature between the initial and final value, $B_i$ and $B_f$, the change is defined as $\Delta S_T=S(T,B_i)-S(T,B_f)$. Usually either $B_i$ or $B_f$  is taken as zero. The magnetocaloric effect is direct if $\Delta S_T>0$. Apart from the isothermal entropy change for finite field variation, a differential measure $-\left(\partial S/\partial B\right)_{T}$ is used to characterize the effect locally (the sign is added to obtain positive quantity for direct magnetocaloric effect, when the entropy decreases with an increase in magnetic field). 

In the following part of the paper the numerical results will be presented and discussed. They include all the above mentioned quantities for V6. The thermodynamic quantities are expressed per mole for convenience.

\section{Numerical results}

In this section, the numerical results concerning the thermodynamic quantities of interest will be presented and discussed. All the calculations were performed using the Wolfram Mathematica software \cite{Mathematica}).

\subsection{Energy levels}

\begin{figure}[ht!]
  \begin{center}
   \includegraphics[width=0.99\columnwidth]{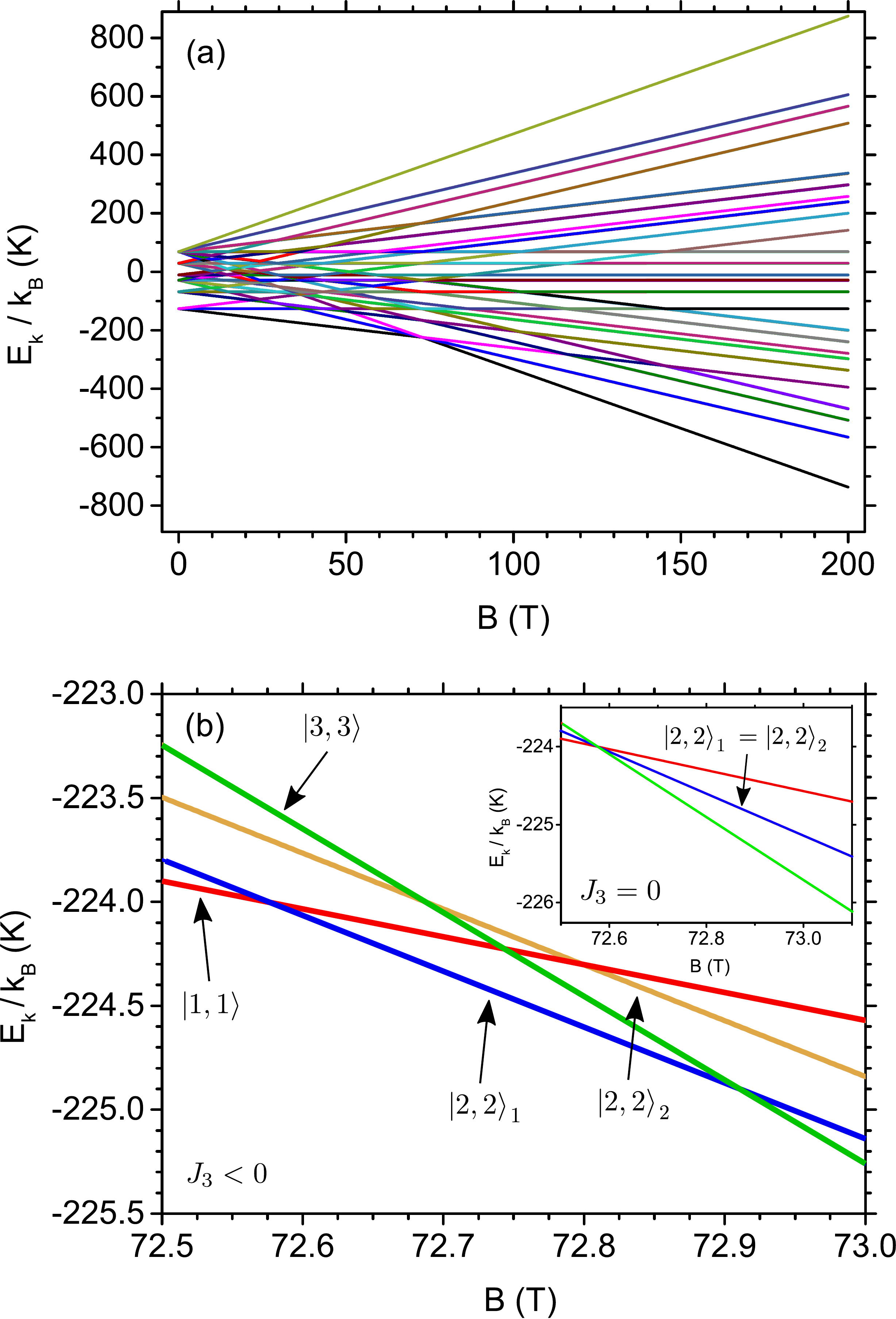}
  \end{center}
   \caption{\label{fig:fige}Energies of a few lowest-lying Hamiltonian eigenstates as a function of the magnetic field (a) in a full range of magnetic fields (b) in the vicinity of the critical fields. In (b), the values of total spin quantum number and spin projection quantum number are indicated. The inset shows the results obtained in the absence of intertriangle coupling. }
\end{figure}

\begin{figure}[ht!]
  \begin{center}
   \includegraphics[width=0.99\columnwidth]{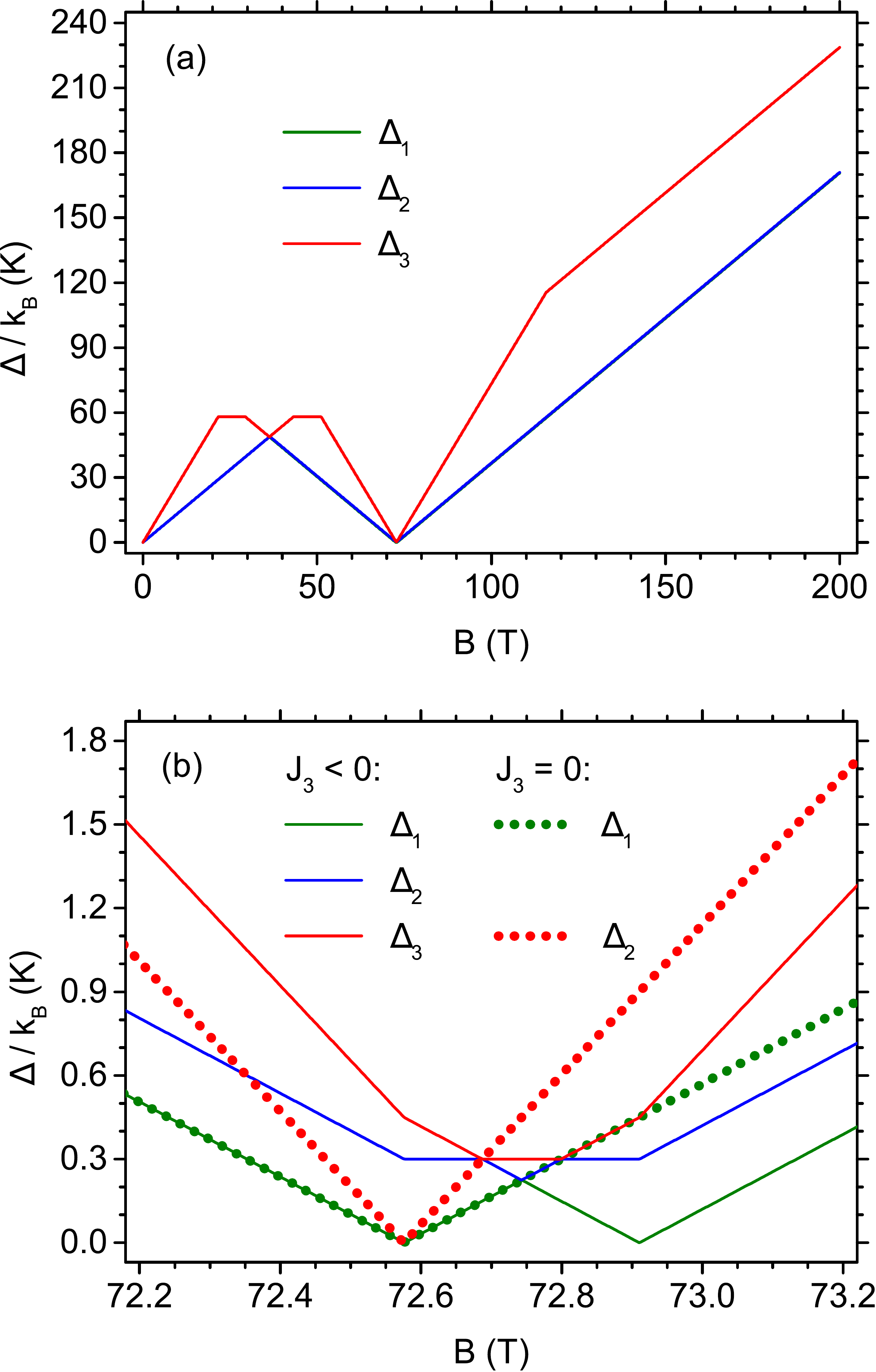}
  \end{center}
   \caption{\label{fig:figgap}Energy gaps between the ground state and the $i$-th lowest-lying excited state as a function of the magnetic field (a) in the full range of magnetic fields and (b) in the vicinity of the critical magnetic fields. In (b), the results in the presence and in the absence of the intertriangle coupling are compared.}
\end{figure}

Let us commence the discussion of the properties of V6 from the analysis of the behaviour of the Hamiltonian eigenstates as a function of the magnetic field applied to the system. The magnetic field dependence of the individual eigenenergies is presented in Fig.~\ref{fig:fige}(a). For convenience, the energy values are divided by $k_{\rm B}$ and expressed in temperature units K. In particular, we will focus the further attention on the ground state. The analytic expressions for the eigenenergies of interest can be found in the \ref{append}. The most useful characterization of the quantum state of V6 cluster involves the total spin quantum number as well as the quantum number of total spin projection. In addition, the analogous quantities for the both individual spin triangles can be mentioned (provided that the given state is an eigenstate of total spin and total spin projection for the triangles). After exact numerical diagonalization of the Hamiltonian (Eq.~\ref{h}) we found that for $B=0$ the ground state is threefold degenerate and corresponds to the state with the total spin $S = 1$ (with projections $S^z=$ 0, $\pm 1$). It should be mentioned that another state, with $S=0$ and $S^z = 0$ is present just above the previous state (their energy difference divided by Boltzmann constant is equal to just 0.136 mK). For finite magnetic fields below the critical field of 72.58 T, the total spin quantum number for the whole V6 cluster is $S = 1$ and the total spin projection quantum number amounts to $S^z = 1$ for the ground state. Moreover, each individual  triangle unit, A and B, is in the state with the total spin $S = 1/2$ and the total spin projection $S^z = 1/2$. On the other side, for fields exceeding the critical field of 72.91 T, the quantum state of the whole cluster is characterized with $S = 3$ and $S^z = 3$. Under such conditions the triangles take the states with $S = 3/2$ and $S^z = 3/2$. An interesting situation emerges in a narrow range of fields, between 72.58 T and 72.91 T. Namely, the V6 cluster is in the state with $S = 2$ and $S^z = 2$. However, both triangles are not in the eigenstates of the squared total spin operator, whereas the spin projection quantum number is $S^z = 1$. It follows that the quantum state of each triangle is a mixed state being an equal statistical mixture of states with $S^z = 1/2$ and $S^z=3/2$. It takes the form of $\rho_{\rm triangle}=\frac{1}{2}\left|-\frac{1}{6},\frac{1}{3},\frac{1}{3}\right>\left<-\frac{1}{6},\frac{1}{3},\frac{1}{3}\right|+\frac{1}{2}\left|\frac{1}{2},\frac{1}{2},\frac{1}{2}\right>\left<\frac{1}{2},\frac{1}{2},\frac{1}{2}\right|$ (where the three subsequent numbers give the projections of the $z$ components of spins located at sites 1,2 and 3 in the triangle). Let us mention here that the case of non-interacting triangles in V6 has been discussed in Ref.~\cite{Luban2002}, where the critical field of 74 T was found to separate the states with spin 1/2 and 3/2 of each individual spin triangle (however, without the intermediate range discussed above by us). The expressions for the critical magnetic fields mentioned above are given in the ~\ref{append}. In Fig.~\ref{fig:fige}(b) the dependence of the energies of four states lowest in energy on the magnetic field can be followed in the vicinity of the critical fields mentioned in the above discussion (the expressions for the eigenenergies and for the critical fields are provided in \ref{append}). The labels give information about the total spin quantum numbers for individual states plotted in Fig.~\ref{fig:fige}(b). The inset presents the results of analogous calculations performed for $J_3 = 0$ (which reproduces the situation discussed in Ref.~\cite{Luban2002}, with a slight difference in the value of the critical field due to difference in exchange integrals accepted). Let us mention that one of the states plotted in the inset in Fig.~\ref{fig:fige}(b) is two-fold degenerate and the degeneracy is lifted by $J_3 \neq 0$ (as seen in the main panel of the figure and discussed in the \ref{append}). Both states mentioned correspond to $S = 2$ and $S^z = 2$ and individual triangles A and B share the same quantum state in both situations. However, both states $\left|2,2\right\rangle_{1,2}$ differ in such a way that the states of spin pairs taken from different triangles are different. It can be verified that correlations between the $x$ components or between the $y$ components of the spins originating from different triangles are opposite in sign, whereas the correlations between z components of spins are identical.
 
To supplement the discussion of the ground state and its dependence on the magnetic field, we present Fig.~\ref{fig:figgap} which shows the analogous dependence of the few energy gaps $\Delta_i$ (where $\Delta_i$ is the energy difference between the $i$-th excited state and the ground state). First, is can be noticed that the gaps $\Delta_1$ and $\Delta_2$ are almost identical, so the corresponding lines overlap in the scale of the plot Fig.~\ref{fig:figgap}(a). The gap $\Delta_3$ is usually significantly larger. Therefore, in wide ranges of the field an approximate two-state description of the system might be useful, as discussed in the \ref{append2}. It is also visible that the gap is a piecewise linear function of $B$, taking one maximum at 36.29 T, and then tending to close above 70 T (exactly at the critical fields of 72.58 T and 72.91 T mentioned above). The details of the behaviour of the gap close to the critical fields are shown in Fig.~\ref{fig:figgap}(b), together with the data obtained for $J_3 = 0$ for reference. In this scale the difference between $\Delta_1$ and $\Delta_2$ for $J_3<0$ is noticeable. It is evident that the gap $\Delta_1$ closes at the critical fields. The dependence of the energy gap on the magnetic field will have consequences for the behaviour of the specific heat of the system in question.

\subsection{Entropy}

\begin{figure}[ht!]
  \begin{center}
   \includegraphics[width=0.99\columnwidth]{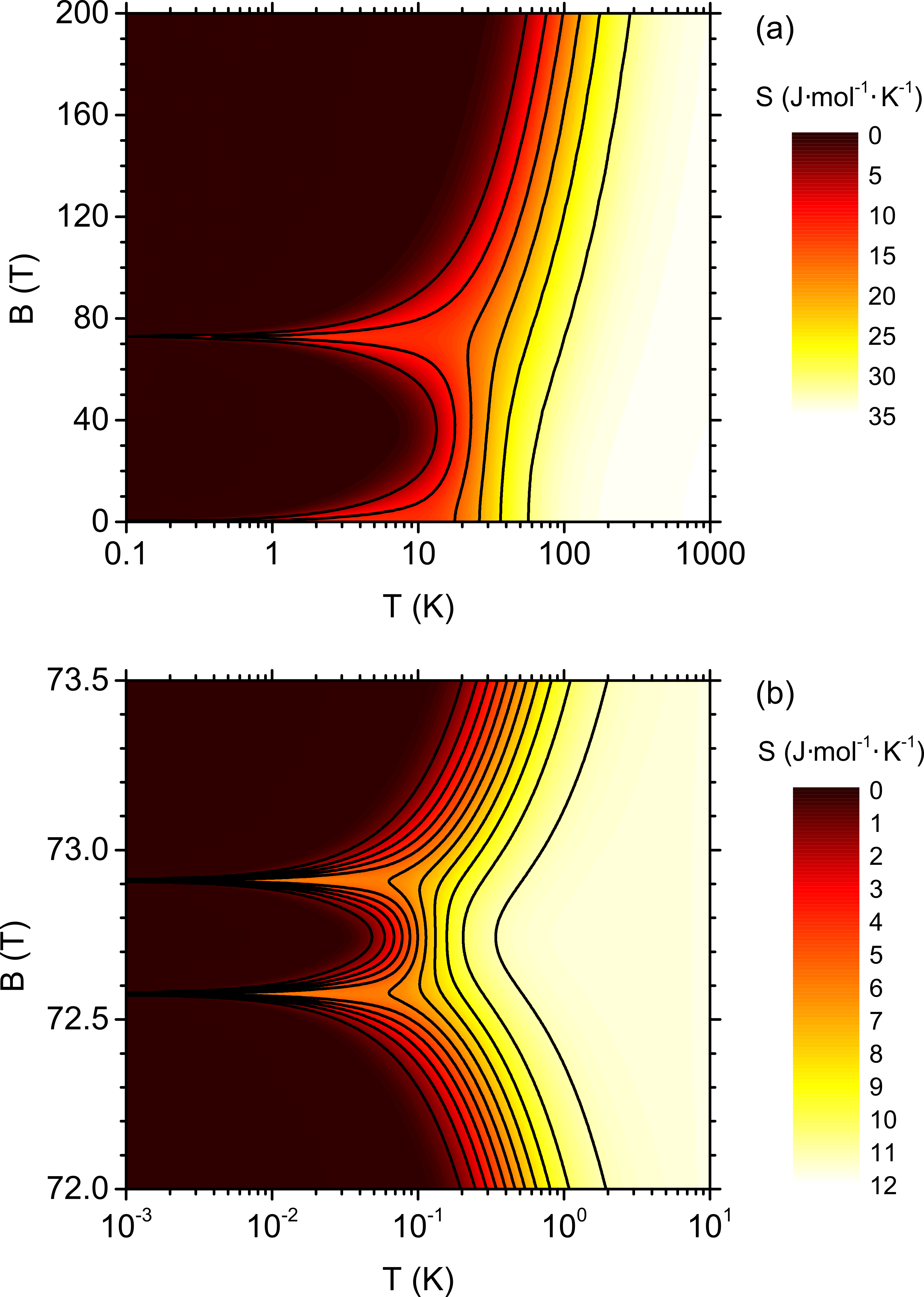}
  \end{center}
   \caption{\label{fig:figS1}Entropy as a function of the temperature and the external magnetic field. Lines of constant entropy are plotted. (a) full range of temperature and magnetic field (b) magnetic fields in the vicinity of critical values.}
\end{figure}
\begin{figure}[ht!]
  \begin{center}
   \includegraphics[width=0.99\columnwidth]{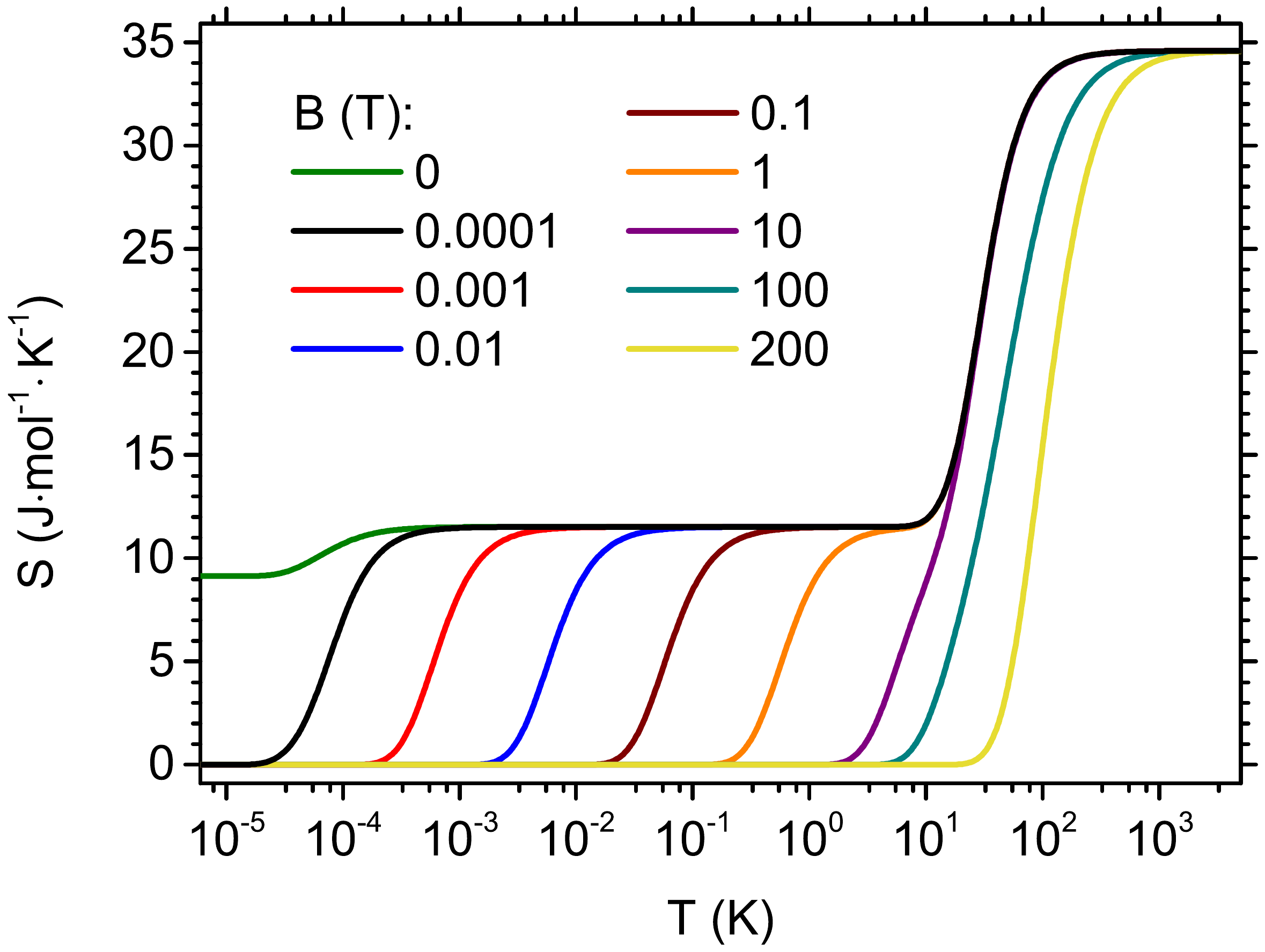}
  \end{center}
   \caption{\label{fig:figS2}Entropy as a function of the temperature for various magnetic fields.}
\end{figure}

The external magnetic field has the principal importance for the magnitude of the magnetocaloric effect, which consists in the dependence of system entropy on the field. Therefore, it is justified to focus firstly on the entropy itself and analyse its variations as a function of the temperature and external magnetic field. The relevant results are collected in two panels in Fig.~\ref{fig:figS1}. The first one presents the entropy variability for a wide range of temperatures (in logarithmic scale) and of magnetic field (linear scale). Significant ranges with rather weak temperature and field dependence of entropy can be noticed. In general, the entropy rises slower with the temperature for higher fields. An interesting behaviour of the entropy can be seen in the vicinity of the field of approximately 70 T, looking like a single pronounced maximum, which is more precisely analysed in the next panel of Fig.~\ref{fig:figS1}. For the temperatures of the order of 1 K we observe a broad maximum and when the temperature is lowered, a sort of ramification occurs, giving rise to somehow finer structure. Therefore, a double peak structure symmetric with respect to some magnetic field builds up; both peaks share the same values of entropy. This behaviour can be traced back to the ground-state behaviour shown in Fig.~\ref{fig:fige}. Close to 72.58 T and 72.91 T an energy gap closes and, as a result, the entropy exhibits peaks. The symmetry comes from the symmetric behaviour of the energy gap (because the ground-state energy is a piecewise linear function of the field). 

The next graph (Fig.~\ref{fig:figS2}) presents the entropy as a function of the temperature; it is different than the previous graphs because we focus on the entropy behaviour for selected values of magnetic field. In particular, at field $B = 0$, the low-temperature residual entropy is seen (equal to $R \ln 3$, as the ground state is three-fold degenerate under these conditions – it is the state with total spin $S = 1$ – and the degeneracy is lifted by $B > 0$). After initial, rather rapid increase at low temperature, the entropy presents a plateau at $R \ln 4$. This is due to the fact that for $B = 0$ the state with $S = 0$ lies considerably close to the threefold degenerate state with $S = 1$. The temperature at which the increase takes place shifts to higher values if the magnetic field increases. As mentioned in the discussion of the energy levels, the two-state description of the thermodynamics is approximately valid for the fields below 36.3 T - see \ref{append2}. On the basis of this model, the entropy can be calculated and the characteristic temperature $T^{**}$ at which the entropy slope $\left(\partial S/\partial T\right)_B$ is maximized can be determined. It follows that $T^{**}=0.29\left(g\mu_{\rm B}/k_{\rm B}\right)B$, giving a proportionality coefficient of 0.39 $K/T$. After the plateau a second increase of entropy to saturation value of $R \ln 64$ is visible. The tendency to shift to higher temperatures under the influence of the magnetic field is very weak, so that the intermediate plateau at $R\ln 4$ becomes narrower for higher fields and eventually vanishes completely. The intermediate plateau is an interesting behaviour of high importance for the description of the magnetocaloric effect in the studied system.

\subsection{Specific heat}

	\begin{figure}[ht!]
  \begin{center}
   \includegraphics[width=0.99\columnwidth]{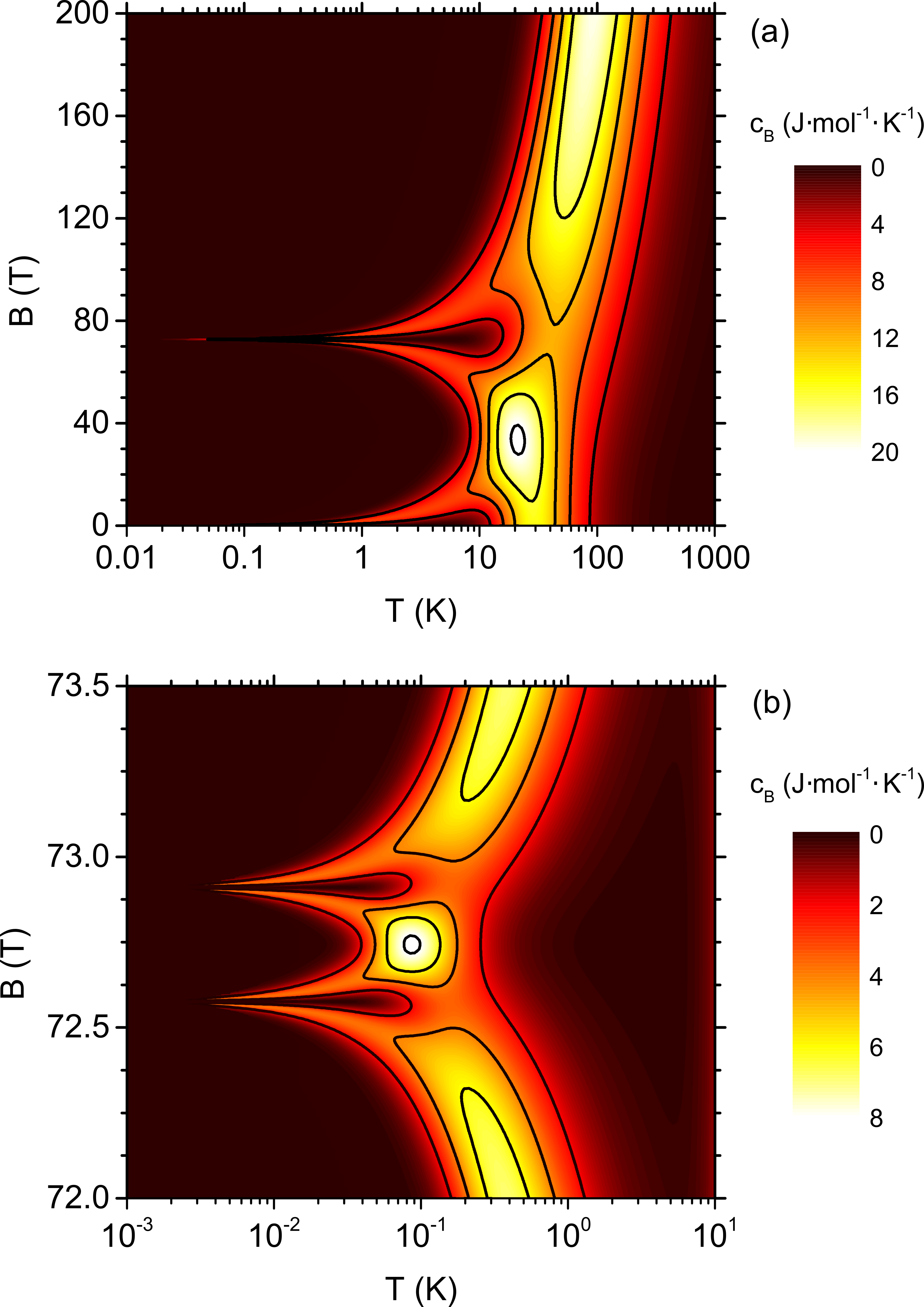}
  \end{center}
   \caption{\label{fig:figC1}Magnetic specific heat as a function of the temperature and the external magnetic field. Lines of constant specific heat are plotted. (a) full range of temperature and magnetic field (b) magnetic fields in the vicinity of critical values.}
\end{figure}
In our calculations we also focused the attention on the specific heat, because it is likewise important aspect for magnetocaloric properties as the entropy. In the same way as previously, we collated together two graphs (Fig.~\ref{fig:figC1}). Fig.~\ref{fig:figC1}(a) shows the specific heat as a function of the temperature and the magnetic field in a wide range of both control parameters. A maximum at intermediate temperatures, the position of which depends on the applied magnetic field, is noticeable. In general, the field shifts the maximum towards higher temperatures, which corresponds to the shift of range in which the entropy slope is large (between the intermediate plateau and the saturation value). This maximum is of Schottky origin and, for low enough fields, appears at the temperature $T^{*}=0.38\left(g\mu_{\rm B}/k_{\rm B}\right)B$ - for details see \ref{append2}. A separate weak maximum at low fields at low temperatures is attributed to the entropy jump between $R \ln 3$ and $R \ln 4$. Like in the case of the entropy, an interesting behaviour is expected close to the critical magnetic fields where the ground state of the system changes. In the panel Fig.~\ref{fig:figC1}(b) a close-up of this range is shown. In the vicinity of both critical magnetic fields a double-peaked structure of specific heat is seen at low temperatures. Then a single peak builds up in between, for intermediate temperatures. However, in general, we observe the largest magnitudes of the specific heat for temperatures between 10 and 100 K.
 
\subsection{Isothermal entropy change}

\begin{figure*}[ht!]
  \begin{center}
   \includegraphics[width=1.98\columnwidth]{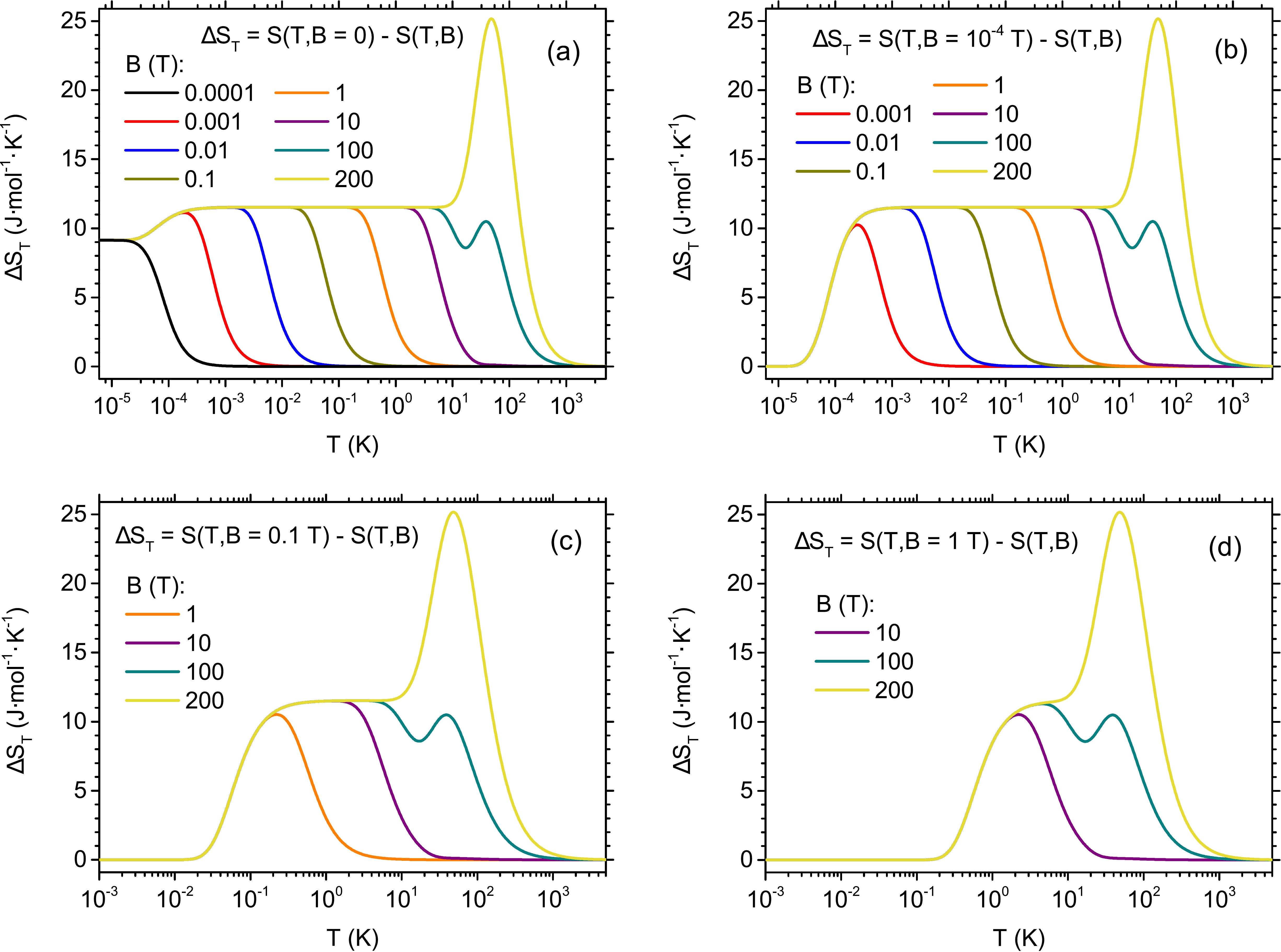}
  \end{center}
   \caption{\label{fig:figds1}Isothermal entropy change as a function of temperature between the initial and final magnetic field, for varying values of final magnetic field: (a) initial field of 0 T; (b) initial field of 10$-4$ T (c) initial field of 0.1 T (d) initial field of 1 T.}
\end{figure*}
\begin{figure}[ht!]
  \begin{center}
   \includegraphics[width=0.99\columnwidth]{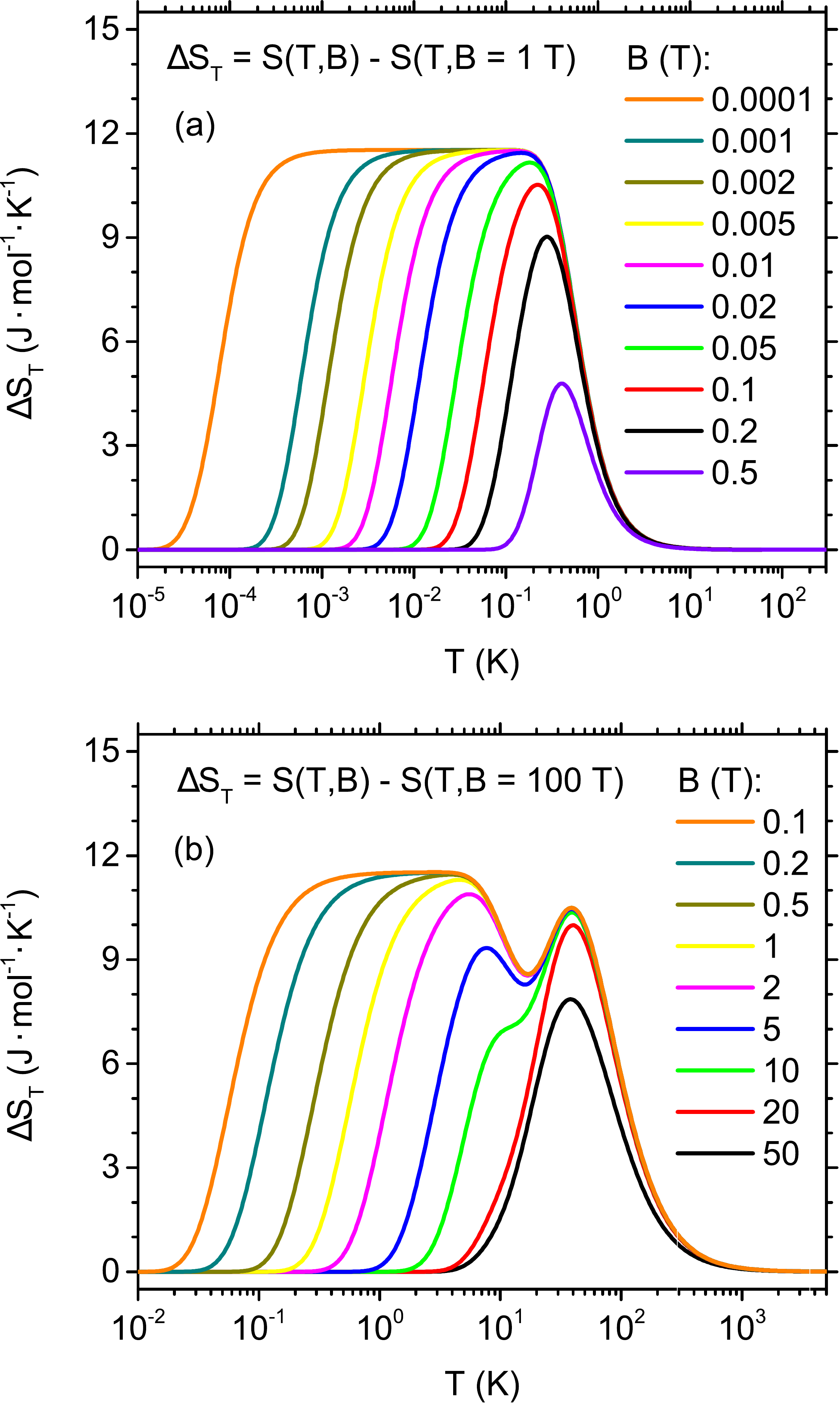}
  \end{center}
   \caption{\label{fig:figds2}Isothermal entropy change as a function of temperature between the initial and final magnetic field, for varying values of initial magnetic field: (a) final field of 1 T; (b) final field of 100 T.}
\end{figure}
\begin{figure}[ht!]
  \begin{center}
   \includegraphics[width=0.99\columnwidth]{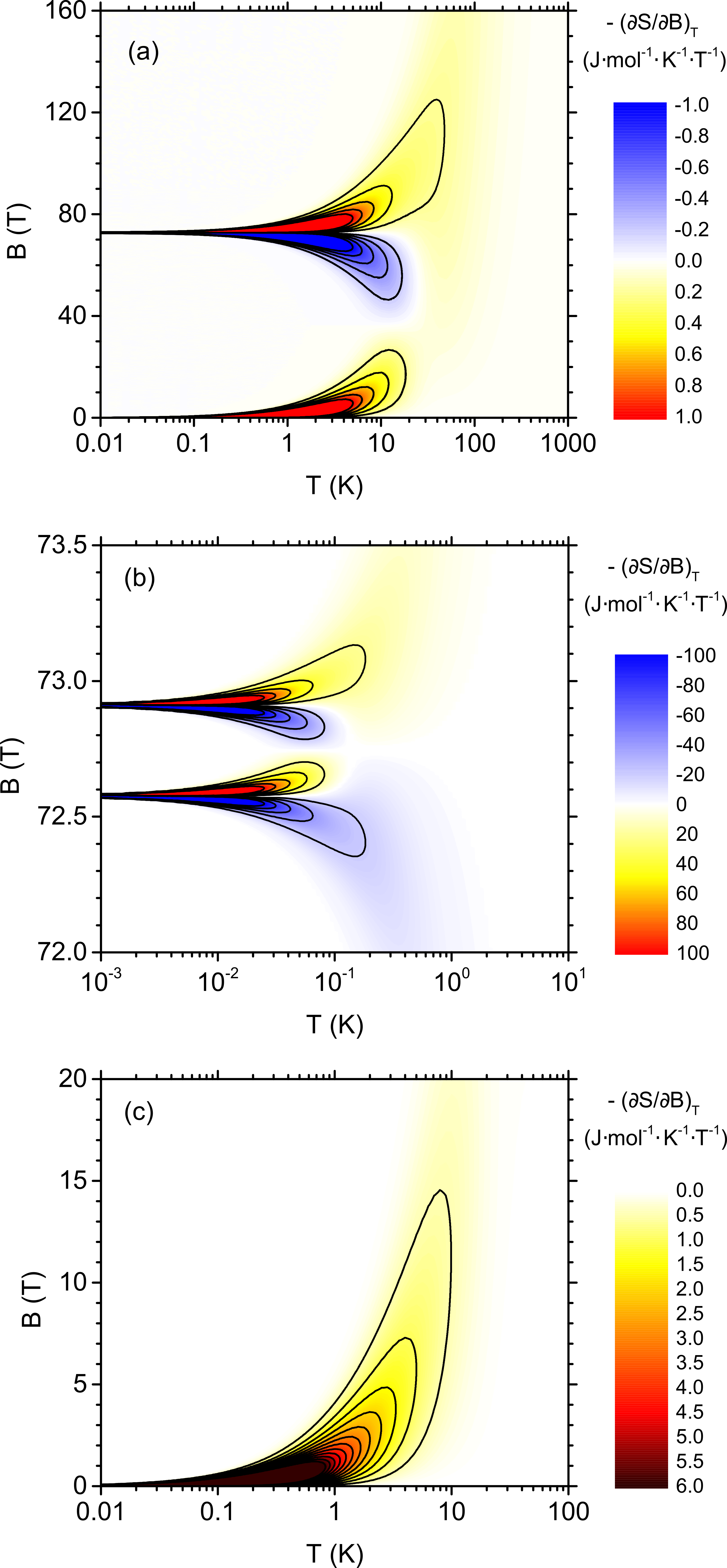}
  \end{center}
   \caption{\label{fig:figdSdB}Differential entropy change, $-\left(\partial S/\partial B\right)_T$, as a function of the temperature and the magnetic field. Lines of constant differential entropy change are plotted. (a) full range of temperature and magnetic field (b) magnetic fields in the vicinity of critical values (c) range of weaker magnetic fields.}
\end{figure}
The crucial quantity characterizing the magnetocaloric effect is the isothermal entropy change, so it deserves primary attention. This quantity is defined as the difference of entropies at two values of magnetic field at constant temperature. Because at $B = 0$ our system indicates nonzero residual entropy of $R \ln 3$, it is reasonable to study the isothermal entropy changes by taking various initial values $B_i$ of the magnetic field, not necessarily zero field. In Fig.~\ref{fig:figds1} the isothermal entropy change is plotted as a function of the temperature for a selection of final values of the magnetic field, up to 200 T. In each of panels (a)-(d) a different initial value of the magnetic field is accepted. 

If the initial field is 0 - Fig.~\ref{fig:figds1}(a) - the entropy change is nonzero even at the lowest field (as $B > 0$ lifts the degeneracy of the ground state). For the low final field the entropy change rather quickly tends to zero. When the final field magnitude increases, the dependence of entropy change on the temperature takes a step-like shape. The characteristic temperature at which the entropy change drops back to zero rises with an increasing final field (and corresponds to the characteristic temperature $T^{**}$ discussed above). At high final fields, exceeding 10 T, a high-temperature maximum emerges. 
 
In Fig.~\ref{fig:figds1}(b), where the low initial field of 10$^{-4}$ T is accepted, the entropy change tends to zero when $T\to 0$. For low final fields, a low-temperature peak gradually builds up. What is crucial, the peak becomes wider as the final field magnitude increases (as the characteristic temperature $T^{**}$ is proportional to the field $B$) and its low-temperature part (where the entropy change rises) is completely insensitive to the final field, thus the whole curve becomes step-like in shape. If the final field exceeds 10 T, an additional maximum builds up at high temperatures and at highest considered final field of 200 T this maximum strongly dominates. It can be summarized that the influence of the increasing final field consists in increasing the width of the step-like curve proportionally to the final field $B_f$. If the higher values of initial field are selected, like 0.1 T in Fig.~\ref{fig:figds1}(c) and 1 T in Fig.~\ref{fig:figds1}(d), the tendency described above is conserved. The only effect is that the initial rise in entropy change is shifted to higher temperatures and the whole plateau of $\Delta S_T$ is narrower.

Due to the highly interesting behaviour of the isothermal entropy change presented in Fig.~\ref{fig:figds1}, we prepared additional Fig.~\ref{fig:figds2}, where analogous quantity is shown but this time the final field magnitude $B_f$ is fixed and the curves are plotted for various values of the initial magnetic field. In Fig.~\ref{fig:figds2}(a) the final field was set to 0.1 T. If the initial field is only slightly smaller, a maximum builds up between 0.1 K and 1 K, and when the initial field is gradually lowered, the peak height increases. Further reduction of the initial magnetic field results in increase of the peak width, as its left side is shifted towards low temperatures (following the changes of $T^{**}$ as a function of $B_i$) and the whole dependence becomes step-like in shape. If the final field is set to 200 T, as shown in Fig.~\ref{fig:figds2}(b), the general behaviour of the isothermal entropy change is analogous. However, this time the maximum emerges initially between 10 K and 100 K; moreover, when the initial field is lowered, the shape of the dependence indicates some dip at about 20 K (while the remaining part of the dependence is step-like in shape and the initial rise of entropy change takes place at lower temperature if the initial field decreases).

In all the cases discussed for Fig.~\ref{fig:figds1} and Fig.~\ref{fig:figds2}, the entropy change value at the plateau of the dependence (so the height of the step) is equal to $R \ln 4$. 

To supplement the results concerning the isothermal entropy changes for finite difference in magnetic fields, also the derivative of the entropy with respect to the magnetic field at fixed temperature, $-\left(\partial S/\partial B\right)_T$, can be studied (the sign has been selected to assure positive values for direct magnetocaloric effect). Such quantity has been plotted in Fig.~\ref{fig:figdSdB}(a) for the full range of parameters. It is visible that the differential isothermal entropy change tends to the significant values (positive or negative) in two ranges. The first range of significant magnitudes of $-\left(\partial S/\partial B\right)_T$ is close to 73 T (so in the vicinity of the critical magnetic fields discussed previously) for temperatures lower than 10 K. The finer structure of the temperature and field dependence of differential entropy change in the vicinity of the critical fields is shown in Fig.~\ref{fig:figdSdB}(b). There the presence of a doubled structure can be proved. Below each critical field the quantity $-\left(\partial S/\partial B\right)_T$ takes pronounced positive values, then it passes zero and just above the critical field the negative values are taken. Therefore, if the field is varied at constant temperature, then a maximum at positive values, the zero value and a minimum at negative values are passed subsequently (in the subkelvin range of temperatures). As a consequence, it is demonstrated that the most pronounced differential changes of entropy occur in the vicinity of the field-induced transitions. Another range of significant magnitudes of $-\left(\partial S/\partial B\right)_T$ corresponds to low magnetic fields and the quantity is maximized at the temperatures of a few K (this is related to rather sharp entropy jump between the values of $R \ln 4$ and 0 in the field), as illustrated in details in Fig.~\ref{fig:figdSdB}(c). 

\section{Final remarks}

In the paper we have studied theoretically the magnetocaloric effect in V6 molecular magnet using an exact approach based on the Heisenberg Hamiltonian and canonical (field) ensemble. Our interest was focused on the isothermal entropy change as a quantitative measure of the magnetocaloric effect. The entropy change was studied both for finite changes of magnetic field and as differential quantity $-\left(\partial S/\partial B\right)_T$. Also, the magnetic entropy itself and magnetic component of the specific heat were discussed. The wide range of temperature and magnetic field was taken into account. 

The discussion of the magnetocaloric properties was started from the analysis of the ground state of the system in question. At low magnetic fields the ground state corresponds to total spin of 1, and in the vicinity of 73 T two subsequent transitions take place, to states with total spin of 2 and 3, accordingly. In the absence of the field the ground state is three-fold degenerate, however, the non-degenerate state next in energy lies very close to the ground state. As a consequence, the ground state entropy is $R \ln 3$ but it quickly jumps to $R \ln 4$ when the temperature increases and then, after a plateau, reaches the saturation value of $R \ln 64$.

The mentioned type of entropy dependence on the temperature gives rise to interesting magnetocaloric behaviour with elongated plateaus of direct magnetocaloric effect, possessing step-like shape. The temperature values at which the entropy change rises or falls down rapidly can be tuned with the initial and final value of the magnetic field as the relevant characteristic temperatures are proportional to the field  - see \ref{append2}. This sort of behaviour can be of practical importance and might possess some potential for applications. It should be emphasized that increase in either initial of final magnetic field over a wide range of values leads to increase of width of the step-like dependence of isothermal entropy change on the temperature, while the height of the step is unchanged. This contrasts with an usual behaviour when rather the maximal magnitude of entropy change rises with the increase in field amplitude. 

Close to the critical magnetic fields interesting details in behaviour of the magnetic specific heat and the differential entropy change can be detected. 

The presented computational study might stimulate experimental characterization of the magnetocaloric properties of V6 polyoxovanadate molecular magnets (as it was characterized in the case of V15 \cite{Fu2015}). In addition, the study could be extended to involve other than magnetic degrees of freedom. 

It might be mentioned finally that a geometry of two weakly coupled triangles bears some resemblance to quasi one-dimensional structures in which spin triangles are stacked (forming a three-leg ladder or triangular tube) \cite{Schnack2004a,Ivanov2010,Alecio2016}. Due to weak intertriangle coupling in V6, the physics is dominated by the behaviour intrinsic to the spin triangles themselves. The structure is unfrustrated due to the huge difference in intratriangle couplings. The case of uncoupled triangles would lead to just a single plateau at 1/3 of the saturation magnetization \cite{Schnack2004a}, whereas the comparable inter- and intratriangle couplings cause the magnetization curve to flatten for infinite system \cite{Schnack2004a}. For our finite system the situation is somehow different, as we deal with weakly coupled triangles (and cluster system), so that both a long magnetization plateau at 1/3 and a narrow plateau at 2/3 is predicted. Note that a discussion of the influence of intramolecular couplings on the magnetic behaviour of molecular systems has been presented in Ref.~\cite{Schnack2016a}.

Also the structure of V6 can be regarded as an example of spin octahedron (see for example Fig.~1 in Ref.~\cite{Karlova2017b}). However, the considered structures involved identical interactions between the spins, whereas in V6 there are three values of couplings, differing by an order of magnitude. In the symmetric octahedron studied theoretically in Ref.~\cite{Karlova2017b} the magnetization plateaus were predicted between the 0 value and the saturation value for magnetization. What is interesting, the plateau at 2/3 was absent only in the case of Ising interaction and present for isotropic heisenberg coupling. In our case, the ground state with $S=2$ is absent if the triangles are uncoupled, but it emerges in a narrow range of magnetic fields if $J_3$ is taken into account (the interactions are isotropic in spin space). This behaviours resemble each other to some extent. Also in Fig.~8 of Ref.~\cite{Karlova2017b} the density plot of entropy as a function of the temperature and field is shown and rapid variability of entropy in the vicinity of two critical magnetic fields is predicted, what bears some analogy to our Fig.~\ref{fig:figS1}.

\section*{Acknowledgements}

This work has been supported by Polish Ministry of Science and Higher Education on a special purpose grant to fund the research and development activities and tasks associated with them, serving the development of young scientists and doctoral students.

\appendix

\section{The ground states and critical magnetic fields}
\label{append}

For the case of V6 molecular magnet, the three states with various total spin $S$ and total spin projection $S^z$ can constitute ground states in various ranges of the external magnetic field (for the values of exchange integrals given in the section \ref{theory}). The energies of these states as well as the critical fields corresponding to cross-overs between them are discussed below. 

The energy of the state with $S=1$ and $S^z=1$ for $J_3=0$ (i.e. in the absence of the intertriangle interactions) equals 
\begin{equation}
E_{1}=2 J_1 - \frac{1}{2}J_2 -  g_e  \mu_{\rm B}B.
\end{equation} 

In the case of $J_3\neq 0$ (in the presence of the intertriangle interactions) it is given by the root of the 4th order equation, which we will not present due to its length. However, since $J_3/J_1\ll 1$, an excellent approximation is provided by a series expansion in the following form: 
\begin{equation}\label{e1}
E_{1}\simeq 2 J_1 - \frac{1}{2}J_2 +\frac{29}{72}\frac{J_3^2}{J_1} -  g_e  \mu_{\rm B}B.
\end{equation}

For $J_3=0$ the state with $S=2$ and $S^z=2$ is twofold degenerate and its energy equals to:
\begin{equation}
E_{2}=\frac{1}{2}J_1 - \frac{1}{2}J_2 - 2 g_e \mu_{\rm B} B.
\end{equation} 

In the presence of $J_3\neq 0$ the degeneracy is lifted and the energies of both resulting   states equal
\begin{equation}
E_{2,1}=\frac{1}{2}J_1 - \frac{1}{2}J_2 - 2 g_e \mu_{\rm B} B
\end{equation} 
and 
\begin{equation}
E_{2,2}=\frac{1}{2}J_1 - \frac{1}{2}J_2 -J_3- 2 g_e \mu_{\rm B} B
\end{equation}
(where the latter state is never a ground state). 

For $J_3=0$ the state with $S=3$ and $S^z=3$ has the energy of:
\begin{equation}
E_3=-J_1 - \frac{1}{2}J_2 - 3 g_e \mu_{\rm B} B,
\end{equation} 
which turns into 
\begin{equation}
E_3=-J_1 - \frac{1}{2}J_2 - \frac{3}{2}J_3 - 3 g_e \mu_{\rm B} B
\end{equation} 
when $J_3$ is included.

If $J_3=0$, the ground state changes directly from $S=1$ to $S=3$ at a (single) critical magnetic field (see also \cite{Luban2002}) equal to:
\begin{equation}
B_{c}=\frac{3}{2} \frac{|J_1|}{g_e \mu_{\rm B}}.
\end{equation} 

In the presence of $J_3\neq 0$ we deal with two critical fields.   

The field $B_{c,1}$ at which the total spin changes from $S=1$ to $S=2$ is given by the condition $E_1=E_{2,1}$ and can be calculated exactly by solving the appropriate 4th order equation (we do not include the formula for its length). However using the series expansion (Eq.~\ref{e1}), we obtain the following expression:
\begin{equation}
B_{c,1}\simeq \frac{3}{2} \frac{|J_1|}{g_e \mu_{\rm B}}\left[ 1+\frac{29}{108}\left(\frac{J_3}{J_1}\right)^2\right].
\end{equation} 
It is noticeable that the field $B_{c,1}$ is a quadratic function of $J_3$ for weak $J_3$.

The field $B_{c,2}$ at which the total spin changes from $S=2$ to $S=3$ is given by the condition $E_{2,1}=E_3$ and is expressed by the formula: 
\begin{equation}
B_{c,2}=\frac{3}{2} \frac{|J_1|}{g_e \mu_{\rm B}}\left(1+\frac{|J_3|}{|J_1|}\right).
\end{equation} 

\section{Thermodynamics of the two-state system: entropy and specific heat}
\label{append2}

Suppose that the spectrum of the system in question can be limited to just two energy states - a ground state with energy equal to $E_0$ and the degeneracy of $g_0$ and the first excited state with the energy $E_0+\Delta_1$ and degeneracy of $g_1$. Then, using the standard thermodynamics with the statistical sum $Z=g_0+g_1\exp\left[\Delta_1/\left(k_{\rm B}T\right)\right]$ we can obtain the expression for the entropy of the system:
\begin{equation}
S=k_{\rm B}\left(\ln Z+\frac{g_1\Delta_1}{ZT}\right)
\end{equation}
as well as for the specific heat: 
\begin{equation}
c=k_{\rm B}g_0g_1\frac{\Delta_1^2\exp\left[\Delta_1/\left(k_{\rm B}T\right)\right]}{Z^2T^2}.
\end{equation}
The specific heat is known to exhibit a Schottky maximum \cite{Karlova2016a}. The position of such maximum $T^*$ is given by the solution of the following transcendental equation:
\begin{eqnarray}
g_0\left(1-2\frac{k_{\rm B}T^*}{\Delta_1}\right)\exp\left(\frac{\Delta_1}{k_{\rm B}T^*}\right)&=&g_1\left(1+2\frac{k_{\rm B}T^*}{\Delta_1}\right).\nonumber\\
\end{eqnarray}

On the other hand, the entropy slope $\left(\partial S/\partial T\right)$ exhibits a local maximum for the temperature $T^{**}$ which can be calculated from the transcendental equation:
\begin{eqnarray}
g_0\left(1-3\frac{k_{\rm B}T^{**}}{\Delta_1}\right)\exp\left(\frac{\Delta_1}{k_{\rm B}T^{**}}\right)&=&g_1\left(1+3\frac{k_{\rm B}T^{**}}{\Delta_1}\right).\nonumber\\
\end{eqnarray}

Note that both maxima do not coincide as the specific heat is defined as $c=T\left(\partial S/\partial T\right)$.

Both characteristic temperatures, $T^{*}$ and $T^{**}$, depend on the relative degeneracy of the ground state and the first excited state. In the case of V6 molecular magnet at low magnetic fields we have a non-degenerate ground state ($g_0=1$) and two excited states (one with $S=1,\,S^z=0$ and one with $S=S^z=0$) separated negligibly in energy from each other, so that we can approximately assume a twofold degeneracy of the first excited state, $g_1=2$. For such case we have $k_{\rm B}T^*/\Delta_1=$0.376696 (see also Ref.~\cite{Karlova2016a}) and $k_{\rm B}T^{**}/\Delta_1=$0.292396.

For the magnetic fields $B$ smaller than $3\left|J_1\right|/\left(4g\mu_{\rm B}\right)$=36.29 T, the energy gap between the ground state and the first excited state is equal to $\Delta_1=g\mu_{\rm B}B$ if we assume $J_3=0$ (note that the correction to the energy of the state $E_1$ is quadratic in $J_3/J_1\ll 1$). As a consequence, the maximum of the entropy slope can be expected at the temperature $T^{**}=0.292396\left(g\mu_{\rm B}/k_{\rm B}\right)B$, proportional to the field $B$. The proportionality coefficient is then equal approximately to 0.39 $K/T$. Also the magnetic specific heat should exhibit a peak at $T^{**}=0.376696\left(g\mu_{\rm B}/k_{\rm B}\right)B$.

Note that the case of Schottky maximum of the specific heat has been extensively discussed in Ref.~\cite{Karlova2016a} in the context of the consequences of the relative degeneracy of the ground and the first excited state.

 \bibliographystyle{elsarticle-num}

\begin{thebibliography}{10}
\expandafter\ifx\csname url\endcsname\relax
  \def\url#1{\texttt{#1}}\fi
\expandafter\ifx\csname urlprefix\endcsname\relax\def\urlprefix{URL }\fi
\expandafter\ifx\csname href\endcsname\relax
  \def\href#1#2{#2} \def\path#1{#1}\fi

\bibitem{Pinkowicz2011}
D.~Pinkowicz, S.~Chor{\c{a}}{\.z}y, O.~Stefa{\'n}czyk, An invitation to
  molecular magnetism, Science Progress 94~(2) (2011) 139--183.
\newblock \href {https://doi.org/10.3184/003685011X13046171774898}
  {\path{doi:10.3184/003685011X13046171774898}}.

\bibitem{Sieklucka2017}
B.~Sieklucka, D.~Pinkowicz (Eds.), Molecular {{Magnetic Materials}}:
  {{Concepts}} and {{Applications}}, {Wiley-VCH}, {Weinheim}, 2017.
\newblock \href {https://doi.org/10.1002/9783527694228}
  {\path{doi:10.1002/9783527694228}}.

\bibitem{Blundell2004}
S.~J. Blundell, F.~L. Pratt, Organic and molecular magnets, Journal of Physics:
  Condensed Matter 16~(24) (2004) R771--R828.
\newblock \href {https://doi.org/10.1088/0953-8984/16/24/R03}
  {\path{doi:10.1088/0953-8984/16/24/R03}}.

\bibitem{Gatteschi2006}
D.~Gatteschi, R.~Sessoli, J.~Villain, Molecular {{Nanomagnets}}, {Oxford
  University Press}, 2006.
\newblock \href {https://doi.org/10.1093/acprof:oso/9780198567530.001.0001}
  {\path{doi:10.1093/acprof:oso/9780198567530.001.0001}}.

\bibitem{Friedman2010}
J.~R. Friedman, M.~P. Sarachik, Single-{{Molecule Nanomagnets}}, Annual Review
  of Condensed Matter Physics 1~(1) (2010) 109--128.
\newblock \href {https://doi.org/10.1146/annurev-conmatphys-070909-104053}
  {\path{doi:10.1146/annurev-conmatphys-070909-104053}}.

\bibitem{Ponomaryov2015}
A.~N. Ponomaryov, N.~Kim, Z.~H. Jang, J.~van Tol, H.-J. Koo, J.~M. Law, B.~J.
  Suh, S.~Yoon, K.~Y. Choi, Spin decoherence processes in the {{S}} = 1/2
  scalene triangular cluster {(Cu}$_3${(OH))}, New Journal of Physics 17~(3)
  (2015) 033042.
\newblock \href {https://doi.org/10.1088/1367-2630/17/3/033042}
  {\path{doi:10.1088/1367-2630/17/3/033042}}.

\bibitem{Furukawa2016}
Y.~Furukawa, Review of {{NMR}} studies of nanoscale molecular magnets composed
  of geometrically frustrated antiferromagnetic triangles, International
  Journal of Nanotechnology 13~(10/11/12) (2016) 845.
\newblock \href {https://doi.org/10.1504/IJNT.2016.080351}
  {\path{doi:10.1504/IJNT.2016.080351}}.

\bibitem{Diep2013}
H.~T. Diep, Frustrated {{Spin Systems}}, 2nd Edition, {World Scientific}, 2013.
\newblock \href {https://doi.org/10.1142/8676} {\path{doi:10.1142/8676}}.

\bibitem{Hu2015}
H.-C. Hu, C.-S. Cao, Y.~Yang, P.~Cheng, B.~Zhao, A triangular [{{Mn}}$_3$]
  cluster-based ferrimagnet with significant magnetic entropy change, Journal
  of Materials Chemistry C 3~(14) (2015) 3494--3499.
\newblock \href {https://doi.org/10.1039/C4TC02958E}
  {\path{doi:10.1039/C4TC02958E}}.

\bibitem{Choi2008}
K.-Y. Choi, N.~S. Dalal, A.~P. Reyes, P.~L. Kuhns, Y.~H. Matsuda, H.~Nojiri,
  S.~S. Mal, U.~Kortz, Pulsed-field magnetization, electron spin resonance, and
  nuclear spin-lattice relaxation in the {{Cu}}$_{3}$ spin triangle, Physical
  Review B 77~(2) (2008) 024406.
\newblock \href {https://doi.org/10.1103/PhysRevB.77.024406}
  {\path{doi:10.1103/PhysRevB.77.024406}}.

\bibitem{Hao2013}
X.~Hao, X.~Wang, C.~Liu, S.~Zhu, Finite-temperature decoherence of spin states
  in a {{Cu}}$_{3}$ single molecular magnet, Journal of Physics B: Atomic,
  Molecular and Optical Physics 46~(2) (2013) 025502.
\newblock \href {https://doi.org/10.1088/0953-4075/46/2/025502}
  {\path{doi:10.1088/0953-4075/46/2/025502}}.

\bibitem{Nath2013}
R.~Nath, A.~A. Tsirlin, P.~Khuntia, O.~Janson, T.~F{\"o}rster, M.~Padmanabhan,
  J.~Li, Y.~Skourski, M.~Baenitz, H.~Rosner, I.~Rousochatzakis, Magnetization
  and spin dynamics of the spin $s=1/2$ hourglass nanomagnet
  {Cu}$_5${(OH)}$_2${(NIPA)}$_4$$\cdot$10{H}$_2${O}, Physical Review B 87~(21)
  (2013) 214417.
\newblock \href {https://doi.org/10.1103/PhysRevB.87.214417}
  {\path{doi:10.1103/PhysRevB.87.214417}}.

\bibitem{Iida2011}
K.~Iida, Y.~Qiu, T.~J. Sato, Dzyaloshinsky-{{Moriya}} interaction and long
  lifetime of the spin state in the {{Cu}}$_{3}$ triangular spin cluster by
  inelastic neutron scattering measurements, Physical Review B 84~(9) (2011)
  094449.
\newblock \href {https://doi.org/10.1103/PhysRevB.84.094449}
  {\path{doi:10.1103/PhysRevB.84.094449}}.

\bibitem{Muller1998}
A.~M{\"u}ller, J.~Meyer, H.~B{\"o}gge, A.~Stammler, A.~Botar, Trinuclear
  {{Fragments}} as {{Nucleation Centres}}: {{New Polyoxoalkoxyvanadates}} by
  ({{Induced}}) {{Self}}-{{Assembly}}, Chemistry \textendash{} A European
  Journal 4~(8) (1998) 1388--1397.
\newblock \href
  {https://doi.org/10.1002/(SICI)1521-3765(19980807)4:8<1388::AID-CHEM1388>3.0.CO;2-1}
  {\path{doi:10.1002/(SICI)1521-3765(19980807)4:8<1388::AID-CHEM1388>3.0.CO;2-1}}.

\bibitem{Iida2009}
K.~Iida, H.~Ishikawa, T.~Yamase, T.~J. Sato, Inelastic {{Neutron Scattering
  Study}} on {{Anisotropic Exchange}} and {{Dzyaloshinsky}}\textendash{{Moriya
  Interactions}} in the $s=1/2$ {{Triangular Spin Cluster}} {V}$_3$, Journal of
  the Physical Society of Japan 78~(11) (2009) 114709--114709.
\newblock \href {https://doi.org/10.1143/jpsj.78.114709}
  {\path{doi:10.1143/jpsj.78.114709}}.

\bibitem{Belinsky2014}
M.~I. Belinsky, Field-dependent spin chirality and frustration in {{V}}$_3$ and
  {{Cu}}$_3$ nanomagnets in transverse magnetic field. 1. {{Correlations}}
  between variable planar spin configurations, vector and scalar chiralities
  and magnetization, Chemical Physics 435 (2014) 62--94.
\newblock \href {https://doi.org/10.1016/j.chemphys.2013.11.012}
  {\path{doi:10.1016/j.chemphys.2013.11.012}}.

\bibitem{Belinsky2016}
M.~I. Belinsky, Spin {Chirality} of {Cu}$_3$ and {V}$_3$ {Nanomagnets}. 1.
  {{Rotation Behavior}} of {{Vector Chirality}}, {Scalar Chirality}, and
  {Magnetization} in the {Rotating Magnetic Field}, {Magnetochiral
  Correlations}, Inorganic Chemistry 55~(9) (2016) 4078--4090.
\newblock \href {https://doi.org/10.1021/acs.inorgchem.5b02202}
  {\path{doi:10.1021/acs.inorgchem.5b02202}}.

\bibitem{Belinsky2016a}
M.~I. Belinsky, Spin {{Chirality}} of {Cu}$_3$ and {V}$_3$ {Nanomagnets}. 2.
  {{Frustration}}, {{Temperature}}, and {{Distortion Dependence}} of {{Spin
  Chiralities}} and {{Magnetization}} in the {{Rotating}} and {{Tilted Magnetic
  Fields}}, Inorganic Chemistry 55~(9) (2016) 4091--4109.
\newblock \href {https://doi.org/10.1021/acs.inorgchem.5b02204}
  {\path{doi:10.1021/acs.inorgchem.5b02204}}.

\bibitem{Corti2011}
M.~Corti, L.~Cattaneo, M.~C. Mozzati, F.~Borsa, Z.~H. Jang, X.~Fang, Ground
  state of the magnetic molecule $\{${V}6$\}$ determined by broadband electron
  spin resonance at low frequency, Journal of Applied Physics 109~(7) (2011)
  07B104.
\newblock \href {https://doi.org/10.1063/1.3545806}
  {\path{doi:10.1063/1.3545806}}.

\bibitem{Luban2002}
M.~Luban, F.~Borsa, S.~Bud'ko, P.~Canfield, S.~Jun, J.~K. Jung,
  P.~K{\"o}gerler, D.~Mentrup, A.~M{\"u}ller, R.~Modler, D.~Procissi, B.~J.
  Suh, M.~Torikachvili, Heisenberg spin triangles in $\{${V}$_6$$\}$-type
  magnetic molecules: {{Experiment}} and theory, Physical Review B 66~(5)
  (2002) 054407.
\newblock \href {https://doi.org/10.1103/PhysRevB.66.054407}
  {\path{doi:10.1103/PhysRevB.66.054407}}.

\bibitem{Rousochatzakis2005}
I.~Rousochatzakis, Y.~Ajiro, H.~Mitamura, P.~K{\"o}gerler, M.~Luban, Hysteresis
  {{Loops}} and {{Adiabatic Landau}}-{{Zener}}-{{St}}{\"u}ckelberg
  {{Transitions}} in the {{Magnetic Molecule}} $\{$V$_6$$\}$, Physical Review
  Letters 94~(14) (2005) 147204.
\newblock \href {https://doi.org/10.1103/PhysRevLett.94.147204}
  {\path{doi:10.1103/PhysRevLett.94.147204}}.

\bibitem{Jung2002}
J.~K. Jung, D.~Procissi, Z.~H. Jang, B.~J. Suh, F.~Borsa, M.~Luban,
  P.~K{\"o}gerler, A.~M{\"u}ller, $^1${H} and $^{23}${Na} nuclear magnetic
  resonance study of {{V6}} magnetic molecular clusters, Journal of Applied
  Physics 91~(10) (2002) 7391--7393.
\newblock \href {https://doi.org/10.1063/1.1448788}
  {\path{doi:10.1063/1.1448788}}.

\bibitem{Haraldsen2009}
J.~T. Haraldsen, T.~Barnes, J.~W. Sinclair, J.~R. Thompson, R.~L. Sacci,
  J.~F.~C. Turner, Magnetic properties of a {{Heisenberg}} coupled-trimer
  molecular magnet: {{General}} results and application to spin-$\frac{1}{2}$
  vanadium clusters, Physical Review B 80~(6) (2009) 064406.
\newblock \href {https://doi.org/10.1103/PhysRevB.80.064406}
  {\path{doi:10.1103/PhysRevB.80.064406}}.

\bibitem{Barbour2006}
A.~Barbour, R.~D. Luttrell, J.~Choi, J.~L. Musfeldt, D.~Zipse, N.~S. Dalal,
  D.~W. Boukhvalov, V.~V. Dobrovitski, M.~I. Katsnelson, A.~I. Lichtenstein,
  B.~N. Harmon, P.~K{\"o}gerler, Understanding the gap in polyoxovanadate
  molecule-based magnets, Physical Review B 74~(1) (2006) 014411.
\newblock \href {https://doi.org/10.1103/PhysRevB.74.014411}
  {\path{doi:10.1103/PhysRevB.74.014411}}.

\bibitem{Fu2015}
Z.~Fu, Y.~Xiao, Y.~Su, Y.~Zheng, P.~K{\"o}gerler, T.~Br{\"u}ckel, Low-lying
  magnetic excitations and magnetocaloric effect of molecular magnet
  {K}$_6${[V}$_{15}${As}$_6$o$_{42}${(H}$_2${O)]}$\cdot$8{H}$_2${O}, EPL
  112~(2) (2015) 27003.
\newblock \href {https://doi.org/10.1209/0295-5075/112/27003}
  {\path{doi:10.1209/0295-5075/112/27003}}.

\bibitem{Chiorescu2000}
I.~Chiorescu, W.~Wernsdorfer, A.~M{\"u}ller, H.~B{\"o}gge, B.~Barbara,
  Butterfly {{Hysteresis Loop}} and {{Dissipative Spin Reversal}} in the
  $s=1/2$, {V}$_{15}$ {{Molecular Complex}}, Physical Review Letters 84~(15)
  (2000) 3454--3457.
\newblock \href {https://doi.org/10.1103/PhysRevLett.84.3454}
  {\path{doi:10.1103/PhysRevLett.84.3454}}.

\bibitem{Chaboussant2002}
G.~Chaboussant, R.~Basler, A.~Sieber, S.~T. Ochsenbein, A.~Desmedt, R.~E.
  Lechner, M.~T.~F. Telling, P.~K{\"o}gerler, A.~M{\"u}ller, H.-U. G{\"u}del,
  Low-energy spin excitations in the molecular magnetic cluster {{V}}$_{15}$,
  EPL (Europhysics Letters) 59~(2) (2002) 291.
\newblock \href {https://doi.org/10.1209/epl/i2002-00240-x}
  {\path{doi:10.1209/epl/i2002-00240-x}}.

\bibitem{Tarantul2010}
A.~Tarantul, B.~Tsukerblat, Magnetic relaxation in {{V}}$_{15}$ cluster:
  {{Direct}} spin-phonon transitions, Inorganica Chimica Acta 363~(15) (2010)
  4361--4367.
\newblock \href {https://doi.org/10.1016/j.ica.2010.07.080}
  {\path{doi:10.1016/j.ica.2010.07.080}}.

\bibitem{Tarantul2011}
A.~Tarantul, B.~Tsukerblat, Direct and two-phonon {{Orbach}}-{{Aminov}} type
  spin-lattice relaxation in molecular magnet {{V}}$_{15}$, Journal of Physics:
  Conference Series 324 (2011) 012007.
\newblock \href {https://doi.org/10.1088/1742-6596/324/1/012007}
  {\path{doi:10.1088/1742-6596/324/1/012007}}.

\bibitem{Furukawa2005}
Y.~Furukawa, Y.~Fujiyoshi, K.~Kumagai, P.~K{\"o}gerler, Spin dynamics and level
  crossing in nanoscale molecular magnet {{V15}} cluster studied by
  $^1${{H}}-{{NMR}}, Polyhedron 24~(16) (2005) 2737--2744.
\newblock \href {https://doi.org/10.1016/j.poly.2005.03.133}
  {\path{doi:10.1016/j.poly.2005.03.133}}.

\bibitem{Furukawa2007}
Y.~Furukawa, Y.~Nishisaka, K.-i. Kumagai, P.~K{\"o}gerler, F.~Borsa, Local spin
  moment configuration in the frustrated $s=1/2$ {{Heisenberg}} triangular
  antiferromagnet {{V15}} determined by {{NMR}}, Physical Review B 75~(22)
  (2007) 220402.
\newblock \href {https://doi.org/10.1103/PhysRevB.75.220402}
  {\path{doi:10.1103/PhysRevB.75.220402}}.

\bibitem{Salman2009}
Z.~Salman, R.~F. Kiefl, K.~H. Chow, W.~A. MacFarlane, T.~A. Keeler, T.~J.
  Parolin, D.~Wang, Local magnetic susceptibility of the muon in the $s=1/2$
  {{V}}$_{15}$ molecular nano-magnet, Physica B: Condensed Matter 404~(5)
  (2009) 626--629.
\newblock \href {https://doi.org/10.1016/j.physb.2008.11.106}
  {\path{doi:10.1016/j.physb.2008.11.106}}.

\bibitem{Kostyuchenko2008}
V.~V. Kostyuchenko, A.~I. Popov, Exchange interactions and spin states in a
  {{V}}$_{15}$ magnetic molecular nanocluster, Journal of Experimental and
  Theoretical Physics 107~(4) (2008) 595--602.
\newblock \href {https://doi.org/10.1134/S1063776108100063}
  {\path{doi:10.1134/S1063776108100063}}.

\bibitem{Konstantinidis2002}
N.~P. Konstantinidis, D.~Coffey, Magnetic anisotropy in the molecular complex
  {{V}}$_{15}$, Physical Review B 66~(17) (2002) 174426.
\newblock \href {https://doi.org/10.1103/PhysRevB.66.174426}
  {\path{doi:10.1103/PhysRevB.66.174426}}.

\bibitem{Wernsdorfer2004}
W.~Wernsdorfer, A.~M{\"u}ller, D.~Mailly, B.~Barbara, Resonant photon
  absorption in the low-spin molecule {{V}}$_{15}$, EPL (Europhysics Letters)
  66~(6) (2004) 861.
\newblock \href {https://doi.org/10.1209/epl/i2004-10033-9}
  {\path{doi:10.1209/epl/i2004-10033-9}}.

\bibitem{Vongtragool2003}
S.~Vongtragool, B.~Gorshunov, A.~A. Mukhin, J.~van Slageren, M.~Dressel,
  A.~M{\"u}ller, High-frequency magnetic spectroscopy on the molecular magnetic
  cluster {{V}}$_{15}$, Physical Chemistry Chemical Physics 5~(13) (2003)
  2778--2782.
\newblock \href {https://doi.org/10.1039/B302244G}
  {\path{doi:10.1039/B302244G}}.

\bibitem{Sakon2005}
T.~Sakon, K.~Koyama, M.~Motokawa, A.~M{\"u}ller, B.~Barbara, Y.~Ajiro, Detailed
  {{ESR Measurements}} for the {{Quantum Molecular Magnet V15}} at
  {{Ultra}}-{{Low Temperatures}}, Progress of Theoretical Physics Supplement
  159 (2005) 302--307.
\newblock \href {https://doi.org/10.1143/PTPS.159.302}
  {\path{doi:10.1143/PTPS.159.302}}.

\bibitem{Gysler2014}
M.~Gysler, C.~Schlegel, T.~Mitra, A.~M{\"u}ller, B.~Krebs, J.~{van Slageren},
  Spin-forbidden transitions in the molecular nanomagnet {{V}}$_{15}$, Physical
  Review B 90~(14) (2014) 144405.
\newblock \href {https://doi.org/10.1103/PhysRevB.90.144405}
  {\path{doi:10.1103/PhysRevB.90.144405}}.

\bibitem{Procissi2006}
D.~Procissi, A.~Lascialfari, E.~Micotti, M.~Bertassi, P.~Carretta, Y.~Furukawa,
  P.~K{\"o}gerler, Low-energy excitations in the $s=1/2$ molecular nanomagnet
  {K}$_6${[V}$^{IV}_{15}${As}$_6$o$_{42}${(H}$_2${O)]}$\cdot$8{H}$_2${O} from
  proton {{NMR}} and $\mu${SR}, Physical Review B 73~(18) (2006) 184417.
\newblock \href {https://doi.org/10.1103/PhysRevB.73.184417}
  {\path{doi:10.1103/PhysRevB.73.184417}}.

\bibitem{Tsukerblat2018}
B.~Tsukerblat, A.~Tarantul, S.~Aldoshin, A.~Palii, Layered polyoxovanadate
  {{V}}$_{15}$: From structure to highly anisotropic magnetism, Journal of
  Coordination Chemistry 71~(11-13) (2018) 2025--2042.
\newblock \href {https://doi.org/10.1080/00958972.2018.1485900}
  {\path{doi:10.1080/00958972.2018.1485900}}.

\bibitem{Furukawa2007a}
Y.~Furukawa, Y.~Nishisaka, K.~Kumagai, P.~K{\"o}gerler, F.~Borsa, Magnetic
  properties of {{Heisenberg}} triangular antiferromagnet {{V15}} cluster
  studied by {{NMR}}, Hyperfine Interactions 176~(1) (2007) 65--68.
\newblock \href {https://doi.org/10.1007/s10751-008-9614-z}
  {\path{doi:10.1007/s10751-008-9614-z}}.

\bibitem{Popov2004}
A.~I. Popov, V.~I. Plis, A.~F. Popkov, A.~K. Zvezdin, Jahn-{{Teller}} effect in
  multi-spin systems with high exchange interaction: {{Ground}} state problem
  for {V}$_{15}$ nanocluster, Physical Review B 69~(10) (2004) 104418.
\newblock \href {https://doi.org/10.1103/PhysRevB.69.104418}
  {\path{doi:10.1103/PhysRevB.69.104418}}.

\bibitem{Franco2018}
V.~Franco, J.~S. Bl{\'a}zquez, J.~J. Ipus, J.~Y. Law, L.~M.
  {Moreno-Ram{\'i}rez}, A.~Conde, Magnetocaloric effect: {{From}} materials
  research to refrigeration devices, Progress in Materials Science 93 (2018)
  112--232.
\newblock \href {https://doi.org/10.1016/j.pmatsci.2017.10.005}
  {\path{doi:10.1016/j.pmatsci.2017.10.005}}.

\bibitem{deOliveira2010a}
N.~A. {de Oliveira}, P.~J. {von Ranke}, Theoretical aspects of the
  magnetocaloric effect, Physics Reports 489~(4) (2010) 89--159.
\newblock \href {https://doi.org/10.1016/j.physrep.2009.12.006}
  {\path{doi:10.1016/j.physrep.2009.12.006}}.

\bibitem{Sessoli2012}
R.~Sessoli, Chilling with {{Magnetic Molecules}}, Angewandte Chemie
  International Edition 51~(1) (2012) 43--45.
\newblock \href {https://doi.org/10.1002/anie.201104448}
  {\path{doi:10.1002/anie.201104448}}.

\bibitem{Evangelisti2006}
M.~Evangelisti, F.~Luis, L.~J. de~Jongh, M.~Affronte, Magnetothermal properties
  of molecule-based materials, Journal of Materials Chemistry 16~(26) (2006)
  2534--2549.
\newblock \href {https://doi.org/10.1039/B603738K}
  {\path{doi:10.1039/B603738K}}.

\bibitem{Zheng2014}
Y.-Z. Zheng, G.-J. Zhou, Z.~Zheng, R.~E.~P. Winpenny, Molecule-based magnetic
  coolers, Chemical Society Reviews 43~(5) (2014) 1462--1475.
\newblock \href {https://doi.org/10.1039/C3CS60337G}
  {\path{doi:10.1039/C3CS60337G}}.

\bibitem{Garlatti2013}
E.~Garlatti, S.~Carretta, J.~Schnack, G.~Amoretti, P.~Santini, Theoretical
  design of molecular nanomagnets for magnetic refrigeration, Applied Physics
  Letters 103~(20) (2013) 202410.
\newblock \href {https://doi.org/10.1063/1.4830002}
  {\path{doi:10.1063/1.4830002}}.

\bibitem{Liu2016g}
J.-L. Liu, Y.-C. Chen, M.-L. Tong, Molecular {{Design}} for {{Cryogenic
  Magnetic Coolants}}, The Chemical Record 16~(2) (2016) 825--834.
\newblock \href {https://doi.org/10.1002/tcr.201500278}
  {\path{doi:10.1002/tcr.201500278}}.

\bibitem{Evangelisti2014}
M.~Evangelisti, Molecule-{{Based Magnetic Coolers}}: {{Measurement}},
  {{Design}} and {{Application}}, in: Molecular {{Magnets}}, {{NanoScience}}
  and {{Technology}}, {Springer, Berlin, Heidelberg}, 2014, pp. 365--387.
\newblock \href {https://doi.org/10.1007/978-3-642-40609-6_14}
  {\path{doi:10.1007/978-3-642-40609-6_14}}.

\bibitem{Evangelisti2010}
M.~Evangelisti, E.~K. Brechin, Recipes for enhanced molecular cooling, Dalton
  Transactions 39~(20) (2010) 4672--4676.
\newblock \href {https://doi.org/10.1039/B926030G}
  {\path{doi:10.1039/B926030G}}.

\bibitem{Liu2014}
J.-L. Liu, Y.-C. Chen, F.-S. Guo, M.-L. Tong, Recent advances in the design of
  magnetic molecules for use as cryogenic magnetic coolants, Coordination
  Chemistry Reviews 281 (2014) 26--49.
\newblock \href {https://doi.org/10.1016/j.ccr.2014.08.013}
  {\path{doi:10.1016/j.ccr.2014.08.013}}.

\bibitem{Sharples2013}
J.~W. Sharples, D.~Collison, Coordination compounds and the magnetocaloric
  effect, Polyhedron 54 (2013) 91--103.
\newblock \href {https://doi.org/10.1016/j.poly.2013.02.034}
  {\path{doi:10.1016/j.poly.2013.02.034}}.

\bibitem{Evangelisti2005}
M.~Evangelisti, A.~Candini, A.~Ghirri, M.~Affronte, E.~K. Brechin, E.~J.~L.
  McInnes, Spin-enhanced magnetocaloric effect in molecular nanomagnets,
  Applied Physics Letters 87~(7) (2005) 072504.
\newblock \href {https://doi.org/10.1063/1.2010604}
  {\path{doi:10.1063/1.2010604}}.

\bibitem{Torres2000}
F.~Torres, J.~M. Hern{\'a}ndez, X.~Bohigas, J.~Tejada, Giant and time-dependent
  magnetocaloric effect in high-spin molecular magnets, Applied Physics Letters
  77~(20) (2000) 3248--3250.
\newblock \href {https://doi.org/10.1063/1.1325393}
  {\path{doi:10.1063/1.1325393}}.

\bibitem{Balanda2016}
M.~Ba{\l}anda, R.~Pe{\l}ka, M.~Fitta, {\L}.~Laskowski, M.~Laskowska, Relaxation
  and magnetocaloric effect in the {{Mn}}${_12}$ molecular nanomagnet
  incorporated into mesoporous silica: A comparative study, RSC Advances 6~(54)
  (2016) 49179--49186.
\newblock \href {https://doi.org/10.1039/C6RA04063B}
  {\path{doi:10.1039/C6RA04063B}}.

\bibitem{Evangelisti2009}
M.~Evangelisti, A.~Candini, M.~Affronte, E.~Pasca, L.~J. {de Jongh}, R.~T.~W.
  Scott, E.~K. Brechin, Magnetocaloric effect in spin-degenerated molecular
  nanomagnets, Physical Review B 79~(10) (2009) 104414.
\newblock \href {https://doi.org/10.1103/PhysRevB.79.104414}
  {\path{doi:10.1103/PhysRevB.79.104414}}.

\bibitem{Pineda2016}
E.~M. Pineda, G.~Lorusso, K.~H. Zangana, E.~Palacios, J.~Schnack,
  M.~Evangelisti, R.~E.~P. Winpenny, E.~J.~L. McInnes, Observation of the
  influence of dipolar and spin frustration effects on the magnetocaloric
  properties of a trigonal prismatic $\{${Gd}$_7$$\}$ molecular nanomagnet,
  Chemical Science 7~(8) (2016) 4891--4895.
\newblock \href {https://doi.org/10.1039/C6SC01415A}
  {\path{doi:10.1039/C6SC01415A}}.

\bibitem{Kortus2001}
J.~Kortus, M.~Pederson, C.~Hellberg, S.~Khanna, {{DFT}} studies of the
  molecular nanomagnet {{Fe}}$_8$ and the {{V}}$_{15}$ spin system, The
  European Physical Journal D - Atomic, Molecular, Optical and Plasma Physics
  16~(1) (2001) 177--180.
\newblock \href {https://doi.org/10.1007/s100530170086}
  {\path{doi:10.1007/s100530170086}}.

\bibitem{Boukhvalov2004}
D.~W. Boukhvalov, V.~V. Dobrovitski, M.~I. Katsnelson, A.~I. Lichtenstein,
  B.~N. Harmon, P.~K{\"o}gerler, Electronic structure and exchange interactions
  in {{V}}$_{15}$ magnetic molecules: {LDA+U} results, Physical Review B 70~(5)
  (2004) 054417.
\newblock \href {https://doi.org/10.1103/PhysRevB.70.054417}
  {\path{doi:10.1103/PhysRevB.70.054417}}.

\bibitem{Kortus2001a}
J.~Kortus, C.~S. Hellberg, M.~R. Pederson, Hamiltonian of the {{V}}$_{15}$
  {{Spin System}} from {{First}}-{{Principles Density}}-{{Functional
  Calculations}}, Physical Review Letters 86~(15) (2001) 3400--3403.
\newblock \href {https://doi.org/10.1103/PhysRevLett.86.3400}
  {\path{doi:10.1103/PhysRevLett.86.3400}}.

\bibitem{Zukovic2014a}
M.~{\v Z}ukovi{\v c}, A.~Bob{\'a}k, Entropy of spin clusters with frustrated
  geometry, Physics Letters A 378~(26) (2014) 1773--1779.
\newblock \href {https://doi.org/10.1016/j.physleta.2014.04.063}
  {\path{doi:10.1016/j.physleta.2014.04.063}}.

\bibitem{Zukovic2015a}
M.~{\v Z}ukovi{\v c}, Thermodynamic and magnetocaloric properties of
  geometrically frustrated {{Ising}} nanoclusters, Journal of Magnetism and
  Magnetic Materials 374 (2015) 22--35.
\newblock \href {https://doi.org/10.1016/j.jmmm.2014.08.017}
  {\path{doi:10.1016/j.jmmm.2014.08.017}}.

\bibitem{Zukovic2018a}
M.~{\v Z}ukovi{\v c}, M.~Semjan, Magnetization process and magnetocaloric
  effect in geometrically frustrated {{Ising}} antiferromagnet and spin ice
  models on a `{{Star}} of {{David}}' nanocluster, Journal of Magnetism and
  Magnetic Materials 451 (2018) 311--318.
\newblock \href {https://doi.org/10.1016/j.jmmm.2017.11.076}
  {\path{doi:10.1016/j.jmmm.2017.11.076}}.

\bibitem{Strecka2015c}
J.~Stre{\v c}ka, K.~Kar{\v l}ov{\'a}, T.~Madaras, Giant magnetocaloric effect,
  magnetization plateaux and jumps of the regular {{Ising}} polyhedra, Physica
  B: Condensed Matter 466-467 (2015) 76--85.
\newblock \href {https://doi.org/10.1016/j.physb.2015.03.031}
  {\path{doi:10.1016/j.physb.2015.03.031}}.

\bibitem{Karlova2017b}
K.~Kar{\v l}ov{\'a}, J.~Stre{\v c}ka, J.~Richter, Enhanced magnetocaloric
  effect in the proximity of magnetization steps and jumps of spin-1/2 {{XXZ
  Heisenberg}} regular polyhedra, Journal of Physics: Condensed Matter 29~(12)
  (2017) 125802.
\newblock \href {https://doi.org/10.1088/1361-648X/aa53ab}
  {\path{doi:10.1088/1361-648X/aa53ab}}.

\bibitem{Roxburgh2018}
A.~Roxburgh, J.~T. Haraldsen, Thermodynamics and spin mapping of quantum
  excitations in a {{Heisenberg}} spin heptamer, Physical Review B 98~(21)
  (2018) 214434.
\newblock \href {https://doi.org/10.1103/PhysRevB.98.214434}
  {\path{doi:10.1103/PhysRevB.98.214434}}.

\bibitem{Mohylna2019a}
M.~Mohylna, M.~{\v Z}ukovi{\v c}, Magnetocaloric properties of frustrated
  tetrahedra-based spin nanoclusters, Physics Letters A 383~(21) (2019)
  2525--2534.
\newblock \href {https://doi.org/10.1016/j.physleta.2019.05.015}
  {\path{doi:10.1016/j.physleta.2019.05.015}}.

\bibitem{Schnack2013}
J.~Schnack, C.~Heesing, Application of the finite-temperature {{Lanczos}}
  method for the evaluation of magnetocaloric properties of large magnetic
  molecules, The European Physical Journal B 86~(2) (2013) 46.
\newblock \href {https://doi.org/10.1140/epjb/e2012-30546-7}
  {\path{doi:10.1140/epjb/e2012-30546-7}}.

\bibitem{Schnack2013b}
J.~Schnack, J.~Ummethum, Advanced quantum methods for the largest magnetic
  molecules, Polyhedron 66 (2013) 28--33.
\newblock \href {https://doi.org/10.1016/j.poly.2013.01.012}
  {\path{doi:10.1016/j.poly.2013.01.012}}.

\bibitem{Raghu2001}
C.~Raghu, I.~Rudra, D.~Sen, S.~Ramasesha, Properties of low-lying states in
  some high-nuclearity {{Mn}}, {{Fe}}, and {{V}} clusters: {{Exact}} studies of
  {{Heisenberg}} models, Physical Review B 64~(6) (2001) 064419.
\newblock \href {https://doi.org/10.1103/PhysRevB.64.064419}
  {\path{doi:10.1103/PhysRevB.64.064419}}.

\bibitem{Plascak2014}
J.~Plascak, Ensemble thermodynamic potentials of magnetic systems, Journal of
  Magnetism and Magnetic Materials 468 (2018) 224 -- 229.
\newblock \href {https://doi.org/https://doi.org/10.1016/j.jmmm.2018.08.014}
  {\path{doi:https://doi.org/10.1016/j.jmmm.2018.08.014}}.

\bibitem{Mathematica}
Mathematica, {Version} 8.0.4, Wolfram Research, Inc., Champaign, Illinios,
  2010.

\bibitem{Schnack2004a}
J.~Schnack, H.~Nojiri, P.~K{\"o}gerler, G.~J.~T. Cooper, L.~Cronin, Magnetic
  characterization of the frustrated three-leg ladder compound {[(Cu Cl}$_2$
  tach {H})$_3${Cl}]{Cl}$_2$, Physical Review B 70~(17) (2004) 174420.
\newblock \href {https://doi.org/10.1103/PhysRevB.70.174420}
  {\path{doi:10.1103/PhysRevB.70.174420}}.

\bibitem{Ivanov2010}
N.~B. Ivanov, J.~Schnack, R.~Schnalle, J.~Richter, P.~K{\"o}gerler, G.~N.
  Newton, L.~Cronin, Y.~Oshima, H.~Nojiri, Heat {{Capacity Reveals}} the
  {{Physics}} of a {{Frustrated Spin Tube}}, Physical Review Letters 105~(3)
  (2010) 037206.
\newblock \href {https://doi.org/10.1103/PhysRevLett.105.037206}
  {\path{doi:10.1103/PhysRevLett.105.037206}}.

\bibitem{Alecio2016}
R.~C. Al{\'e}cio, M.~L. Lyra, J.~Stre{\v c}ka, Ground states, magnetization
  plateaus and bipartite entanglement of frustrated spin-1/2
  {{Ising}}-{{Heisenberg}} and {{Heisenberg}} triangular tubes, Journal of
  Magnetism and Magnetic Materials 417 (2016) 294--301.
\newblock \href {https://doi.org/10.1016/j.jmmm.2016.05.081}
  {\path{doi:10.1016/j.jmmm.2016.05.081}}.

\bibitem{Schnack2016a}
J.~Schnack, Influence of intermolecular interactions on magnetic observables,
  Physical Review B 93~(5) (2016) 054421.
\newblock \href {https://doi.org/10.1103/PhysRevB.93.054421}
  {\path{doi:10.1103/PhysRevB.93.054421}}.

\bibitem{Karlova2016a}
K.~Kar{\v l}ov{\'a}, J.~Stre{\v c}ka, T.~Madaras, The {{Schottky}}-type
  specific heat as an indicator of relative degeneracy between ground and
  first-excited states: {{The}} case study of regular {{Ising}} polyhedra,
  Physica B: Condensed Matter 488 (2016) 49--56.
\newblock \href {https://doi.org/10.1016/j.physb.2016.01.033}
  {\path{doi:10.1016/j.physb.2016.01.033}}.

\end{thebibliography}


\end{document}